%% file: main.tex
\documentclass[journal,onecolumn]{IEEEtran}
\newcommand{\descr}[1]{\smallskip \noindent \textbf{#1}}
\newcommand{\descrit}[1]{\smallskip \noindent \textbf{\em #1}}

\usepackage[algo2e,ruled]{algorithm2e}
\usepackage{color}
\usepackage{hyperref}
\usepackage[subtle]{savetrees}
\usepackage{xspace}
\usepackage{enumerate}
\usepackage{booktabs}
\usepackage[export]{adjustbox}
\usepackage{multicol}
\usepackage[table,xcdraw]{xcolor}
\usepackage{amssymb}
\usepackage{amsmath}
\usepackage{mathtools}
\DeclarePairedDelimiter\ceil{\lceil}{\rceil}

\allowdisplaybreaks 
\usepackage{amsthm} 
\usepackage{graphicx}
\usepackage[hang,flushmargin]{footmisc}
\usepackage{lipsum}
\usepackage{comment}
\usepackage{xcolor}
\usepackage{alltt}
\usepackage{algorithm}
\usepackage{algorithmic}
\usepackage{setspace}
\usepackage{caption}
\usepackage{subcaption}
\makeatletter
\def\hlinewd#1{%
  \noalign{\ifnum0=`}\fi\hrule \@height #1 \futurelet
   \reserved@a\@xhline}
\makeatother
\usepackage{tabularx}
\usepackage[T1]{fontenc}
\usepackage[utf8]{inputenc}
\usepackage{bm}
\DeclareCaptionType{copyrightbox}

\newtheorem{lemma}{Lemma}
\newtheorem{proposition}{Proposition}

\makeatletter
\newenvironment{protocol}[1][H]{%
    \renewcommand{\ALG@name}{Protocol}
    \begin{algorithm}[#1]%
    }{\end{algorithm}
}
\newcommand{\sbline}{\\[.5\normalbaselineskip]}
\newcommand{\sblinesmall}{\\[.2\normalbaselineskip]}

\newcommand{\sys}{\textsc{spindle}\xspace}
\newcommand{\DHE}{Multiparty Homomorphic Encryption}

\usepackage{newfloat}
\DeclareFloatingEnvironment[
    fileext=los,
    listname=List of Schemes,
    name=Scheme,
    placement=H,
    within=section,
]{scheme}

\newif\ifcomment
\newif\ifminor
\commentfalse
\minorfalse
    
\definecolor{cerise}{rgb}{0.87, 0.19, 0.39}
\definecolor{cadmiumgreen}{rgb}{0.0, 0.42, 0.24}

\ifcomment
	\newcommand{\ap}[1]{\textbf{\em\color{blue}[AP: #1]}}
	\newcommand{\df}[1]{{\color{orange}[DF: #1]}}
	\newcommand{\sinem}[1]{\textbf{\em\color{magenta}[SS: #1]}}
	\newcommand{\jr}[1]{\textbf{\em\color{cadmiumgreen}[JR: #1]}}
	\newcommand{\jpb}[1]{\textbf{\em\color{cerise}[JPB: #1]}}
	\newcommand{\pl}[1]{{\color{gray}[RM?: #1]}}
\else  
    \newcommand\ap[1]{}
    \newcommand\df[1]{#1}
    \newcommand\sinem[1]{}
    \newcommand\jr[1]{}
    \newcommand\jpb[1]{}
    \newcommand{\pl}[1]{#1}
\fi

\ifminor
	\newcommand{\apm}[1]{\textbf{\em\color{blue}[AP: #1]}}
	\newcommand{\dfm}[1]{{\color{blue}#1}}
	\newcommand{\sinemm}[1]{\textbf{\em\color{magenta}[SS: #1]}}
	\newcommand{\jrm}[1]{\textbf{\em\color{cadmiumgreen}[JR: #1]}}
	\newcommand{\jpbm}[1]{\textbf{\em\color{cerise}[JPB: #1]}}
	
\else  
    \newcommand\apm[1]{}
    \newcommand\dfm[1]{#1}
    
    \newcommand\sinemm[1]{}
    \newcommand\jrm[1]{}
    \newcommand\jpbm[1]{}
\fi

\begin{document}
%
\title{Scalable Privacy-Preserving Distributed Learning}

\author{David Froelicher,
Juan R. Troncoso-Pastoriza,
Apostolos Pyrgelis, 
Sinem Sav,\\ Joao Sa Sousa, Jean-Philippe Bossuat and
Jean-Pierre Hubaux
\thanks{This work was partially supported by the grant \#2017-201 of the Strategic Focal Area “Personalized Health and Related Technologies (PHRT)” of the ETH Domain.}
\thanks{D. Froelicher is with the Laboratory for Data Security and DeDiS Laboratory, École Polytechnique Fédérale de Lausanne, 1015 Lausanne, Switzerland, e-mail: david.froelicher@epfl.ch. 
	Juan R. Troncoso-Pastoriza, Apostolos Pyrgelis, Sinem Sav, Joao Sa Sousa, Jean-Philippe Bossuat and Jean-Pierre Hubaux are with the Laboratory for Data Security, École Polytechnique Fédérale de Lausanne, 1015 Lausanne, Switzerland, e-mail: name.surname@epfl.ch.}
}


%



\maketitle

\begin{abstract}
\df{In this paper, we address the problem of privacy-preserving distributed learning and the evaluation of machine-learning models by analyzing it in the widespread MapReduce abstraction that we extend with privacy constraints. We design \sys (Scalable Privacy-preservINg Distributed LEarning), the first distributed and privacy-preserving system that covers the complete ML workflow by enabling the execution of a cooperative gradient-descent and the evaluation of the obtained model and by preserving data and model confidentiality in a passive-adversary model with up to $N-1$ colluding parties. \sys uses multiparty homomorphic encryption to execute parallel high-depth computations on encrypted data without significant overhead. We instantiate \sys for the training and evaluation of generalized linear models on distributed datasets 
and show that it is able to accurately (on par with non-secure centrally-trained models) and efficiently (due to a multi-level parallelization of the computations) train models that require a high number of iterations on large input data with thousands of features, distributed among hundreds of data providers. For instance, it trains a logistic-regression model on a dataset of one million samples with 32 features distributed among 160 data providers in less than three minutes.}
\end{abstract}



\input{introduction}
\input{related}
\input{system}
\input{optimizations}
\input{security2}
\input{evaluation}

\input{discussion}
\input{conclusion}

\bibliographystyle{abbrv}
\bibliography{bibfile_short}

\input{appendix}

\end{document}

%% file: introduction.tex
\section{Introduction}\label{intro}

The training of machine-learning (ML) models usually requires large and diverse datasets \cite{Zhu2016}. In many domains, such as medicine and finance, assembling sufficiently large datasets has proven difficult~\cite{Zhang_BigDataSecurity} and often requires the sharing of data among multiple data-providers. \pl{This is particularly true in medicine, where patients' data are spread among multiple entities: For example, for rare diseases, one hospital might have only a few patients, whereas a medical study requires hundreds of them to produce significant results. Data sharing among many entities, which can be located in multiple countries, is hence required. However, when the data are sensitive and/or personal, they are particularly difficult to share. Data sharing is highly restricted by legal regulations, such as GDPR~\cite{GDPR} in Europe. The financial and reputational consequences of a data breach often make the risk of data sharing higher than its potential benefits.} Hence, it is often impossible to obtain sufficient data to train ML models that are key enablers in medical research~\cite{BigDataMedical}, financial analysis~\cite{BigDataIssues}, and many other domains.

To address this issue, privacy-preserving solutions are gaining interest as they can be key-enablers for ML with sensitive data. Many solutions have been proposed for secure predictions that use pre-trained models~\cite{boemer2019ngraph, bost2015machine,gilad2016cryptonets, juvekar2018gazelle, liu2017oblivious,riazi2019xonn, riazi2018chameleon, rouhani2018deepsecure}. However, the secure training of ML models, which is much more computationally demanding, has been less studied. Some centralized solutions~\cite{aono2016scalable, bonte2018privacy, chen2018logistic, crawford2018doing, graepel2012ml, jiang2019securelr, kim2018logistic, kim2018secure} that rely on homomorphic encryption (HE) were proposed. They have the advantage of being straightforward to implement but require individual records to be transferred out of the control of their owners, which is often not possible, e.g., due to data protection legislation \cite{ndas1, ndas2}. Also, the data are moved to a central repository, which can become a single point of failure. Secure multiparty computation solutions (SMC) proposed for this scenario~\cite{Akavia_WAHC, corrigan2017prio, gascon2017privacy, giacomelli2018privacy,Jagadeesh692, mohassel2017secureml, nikolaenko2013privacy}, \df{often assume that a limited number of computing parties are honest-but-curious and non-colluding}. These assumptions might not hold when the data are sensitive and/or when the parties have competing interests. 
In contrast, homomorphic encryption-based (HE) or hybrid (HE and SMC) solutions~\cite{Drynx, zheng2019helen} that assume a malicious threat model (e.g., Anytrust model~\cite{Wolinsky2012ScalableAG}) focus on limited ML operations (e.g., the training of regularized linear models with low number of features) and are not quantum secure. Recent advances in quantum computing~\cite{IBMQuantum, Quantum1, Quantum4, Quantum3, Quantum2_GoogleAI} have made this technology a potential threat, in a not-so-far future, for existing cryptographic solutions~\cite{Mosca_Quantum}. Google recently announced that they have reached "quantum-supremacy"~\cite{gooquantumsuprem}. \dfm{Even though quantum computers are still far from being able to break state-of-the-art cryptoschemes, we note that certain data (e.g., genomics) remain sensitive over a long period and will be at risk in the future.}

\dfm{\df{Finally, \textit{federated learning}, a non-cryptographic approach for privacy-preserving training of ML models, has recently gained interest. The data remain under the control of their owners and a server coordinates the training by sending the model directly to the data owners, which then update the model with their data. The updated models from multiple participants are averaged to obtain the global model~\cite{federatedLearning1,Konency2016fed}. Recent works have shown that sharing intermediate models with a coordinating server, or among the participants, can lead to various privacy attacks, e.g., extracting participants' inputs~\cite{hitaj2017deep,Wang2019,NIPS2019_9617} or membership inference~\cite{Melis2019,Nasr2019}. To address these problems, multiple works~\cite{Nvidia_Fed,shokri2015privacy,McMahan2018} rely on a differentially private mechanism to obfuscate the intermediate values. However, this obfuscation decreases the data and model utility, whereas the training of accurate models requires high privacy budgets and the achieved privacy level remains unclear~\cite{jayaraman2019evaluating}.}}

\dfm{Existing cryptographic distributed solutions are practical with only a small number of parties and most of the aforementioned solutions focus either on training or on prediction. They neither consider the complete ML workflow nor enable the training of a model that remains secret and enables oblivious prediction on confidential data. In many cases, the trained model is as sensitive as the data on which it is trained, and the use of the model after the training has to be tightly controlled. ML is used in very competitive domains and a balance has to be found between collaboration and competition~\cite{BigDataMedical,AI_competition,BigDataIssues}. For example, entities that collaborate to train a ML model should equally benefit from the resulting model.}

In this paper, we address the problem of privacy-preserving learning and prediction among multiple parties, i.e., data providers (DPs), that do not trust each other. To address this issue, we design a solution that uses the MapReduce abstraction~\cite{dean2008mapreduce} that is often used to define distributed ML tasks~\cite{chu2007map, verbraeken2019survey}. MapReduce defines parallel and distributed algorithms in a simple and well-known abstraction: \textsc{prepare} (data preparation), \textsc{map} (distributed computations executed independently by multiple nodes or machines), \textsc{combine} (combination of the \textsc{map} results, e.g., aggregation) and \textsc{reduce} (computation on the combined results). We build on and extend this abstraction to determine and delimit which information, e.g., \textsc{map} outputs, have to be protected to design a decentralized \textbf{privacy-preserving} system for ML training and prediction. The model is locally trained by the DPs (\textsc{map}) and the results are iteratively combined (\textsc{combine}) to update the global model (\textsc{reduce}). We exploit the partitioned (distributed) data to enable DPs to keep control of their respective data, and we distribute the computation to provide an efficient solution for the training of ML models on confidential data. After the training, the model is kept secret from all entities and is obliviously and collectively used to provide predictions on confidential data that are known only to the entity requesting the prediction. \dfm{We remark that differential-privacy-based federated-learning solutions~\cite{abadi2016deep, chaudhuri2009privacy, du2018privacy, huang2019dp, Nvidia_Fed, shokri2015privacy, kim2020,pathak2010multiparty, jayaraman2018distributed,truex2019hybrid} follow the same model, i.e., they can be defined according to the MapReduce abstraction. However, most solutions introduce a trade-off between accuracy and privacy~\cite{jayaraman2019evaluating}, and do not provide data and model confidentiality simultaneously. On the contrary, our solution uses
a different paradigm in which, similarly to non-secure solutions, the accuracy is traded off with the performance (e.g., number of iterations), but not with privacy.}

We propose \sys (Scalable Privacy-preservINg Distributed LEarning), a system that enables the privacy-preserving\df{, distributed (cooperative) execution of the widespread stochastic mini-batch gradient-descent (SGD) on data that are stored and controlled by multiple DPs.} \sys builds on a state-of-the-art multiparty, lattice-based, quantum-resistant cryptographic scheme to ensure data and model confidentiality, \df{in the passive-adversary model in which all-but-one DPs can be dishonest}. \df{\sys is meant to be a generic and widely-applicable system that supports the SGD-based training of many different ML models. This includes, but is not limited to, support vector machines, graphical models, generalized linear-models and neural networks~\cite{du2018gradient, NeuralBook, Kumar_2015, toulis2014statistical, Zhang_2004}. For concreteness and comparison with existing works, we instantiate \sys for the training of and prediction on generalized linear models (GLMs)~\cite{GLM}, (e.g., linear, logistic and multinomial logistic regressions). GLMs are easily interpretable, capture complex non-linear relations (e.g., logistic regression), and are widely-used in many domains such as finance, engineering, environmental studies and healthcare~\cite{glm_applications}.}


In a realistic scenario where a dataset of 11,500 samples and 90 features is distributed among 10 DPs, \sys efficiently trains a logistic regression model in less than 54 seconds, achieving an accuracy of 83.9\%, equivalent to a non-secure centralized solution. The distribution of the workload enables \sys to efficiently cope with a large number of DPs (parties), as its execution time is practically independent of it. \sys handles a large number of features, by optimizing the use of the cryptosystem's packing capabilities, and by exploiting \textit{single-instruction multiple-data (SIMD)} operations. It is able to perform demanding training tasks, with high number of iterations and thus high-depth computations, by relying on the multiparty cryptoscheme's ability to collectively refresh a ciphertext with no significant overhead. \df{As shown by our evaluation, these properties enable \sys to support training on large and complex data such as imaging or medical datasets. Moreover, \sys scalability over multiple dimensions (features, DPs, data)} represents a notable improvement with respect to state-of-the-art \df{secure solutions ~\cite{zheng2019helen, Drynx}}. 

\noindent In this work, we make the following contributions:
\begin{itemize}
    \item We analyze the problem of privacy-preserving distributed training and of the evaluation of ML models by extending the widespread MapReduce abstraction with privacy constraints. Following this abstraction, we instantiate \sys, \df{the first operational and efficient distributed system that enables the privacy-preserving execution of a complete machine-learning workflow through the use of a cooperative gradient descent on a dataset distributed among many data providers.}
    \item We propose multiple optimizations that enable the efficient use of a quantum-resistant multiparty (N-party) cryptographic scheme by relying on parallel computations, SIMD operations, efficient collective operations and optimized polynomial approximations of the models' activation functions, e.g., sigmoid and softmax.
    \item We propose a method for the parameterization of \sys by capturing the relations among the security and the learning parameters in a graphical model.
    \item We evaluate \sys against centralized and decentralized secure solutions and demonstrate its scalability and accuracy.
\end{itemize}
To the best of our knowledge, \sys is the first operational system that provides the aforementioned features and security guarantees. 

%% file: related.tex
\section{Related Work}\label{sec:related}

\descr{Privacy-Preserving Training of Machine Learning Models.} Some works have focused on \textit{securely outsourcing} the training of linear ML models to the cloud, typically by using homomorphic encryption (HE) techniques~\cite{aono2016scalable, bonte2018privacy, crawford2018doing, graepel2012ml, kim2018logistic, kim2018secure, 8241854}. For instance, Graepel et al.~\cite{graepel2012ml} outsource the training of a linear classifier by employing somewhat HE~\cite{fan2012somewhat}, whereas Aono et al.~\cite{aono2016scalable} approximate logistic regression, and outsource its computation to the cloud by using additive HE~\cite{paillier1999public}. Jiang et al.~\cite{jiang2019securelr} present a framework for outsourcing logistic regression training to public clouds by combining HE with hardware-based security techniques (i.e., Software Guard Extensions). \df{In \sys, we consider a substantially different setting where the sensitive data are distributed among multiple (untrusted) data providers.} 

Along the research direction of \textit{privacy-preserving distributed learning}, most works operate on the two-server model, where data owners encrypt or secret-share their data among two non-colluding servers that are responsible for the computations. For instance, Nikolaenko et al.~\cite{nikolaenko2013privacy} combine additive homomorphic encryption (AHE) and Yao's garbled circuits~\cite{yao1986generate} to enable ridge regression on data that are horizontally partitioned among multiple data providers. Gascon et al.~\cite{gascon2017privacy} extend Nikolaenko et al. work ~\cite{nikolaenko2013privacy} to the case of vertically partitioned datasets and improve its computation time by employing a novel conjugate gradient descend (GD) method, whereas Giacomelli et al.~\cite{giacomelli2018privacy} further reduce computation and communication overheads by using only AHE. Akavia et al.~\cite{Akavia_WAHC} improve the performance of Giacomelli et al. protocols \cite{giacomelli2018privacy} by performing linear regression on packed encrypted data. Mohassel and Zhang~\cite{mohassel2017secureml} develop techniques to handle secure arithmetic operations on decimal numbers, and employ stochastic GD, which, along with multi-party-computation-friendly alternatives for non-linear activation functions, supports the training of logistic regression and neural network models. Schoppmann et al.~\cite{schoppmann2019make} propose data structures that exploit data sparsity to develop secure computation protocols for nearest neighbors, naive Bayes, and logistic regression classification. \sys differs from these approaches as it does not restrict to the two non-colluding server model, and focuses instead on N-party systems, with N$\geqslant$$2$.

Other distributed and privacy-preserving ML approaches employ a three-server model and rely on secret-sharing techniques to train linear regressions~\cite{bogdanov2016rmind}, logistic regressions~\cite{Cho_GWAS}, and neural networks~\cite{mohassel2018aby,wagh2019securenn}. However, such solutions are tailored to the three-party server model and assume an honest majority among the computing parties. An honest majority is also required in the recent work of Rachuri and Suresh~\cite{rachuri2019trident}, who improve on Mohassel and Rindal~\cite{mohassel2018aby} performance by extending its techniques to the four-party setting. Other works focus on the training of ML models among N-parties (N $\geqslant4$), with stronger security assumptions, i.e., each party trusting itself. For instance, Corrigan-Gibbs and Boneh \cite{corrigan2017prio} present Prio, which relies on secret-sharing to enable the training of linear models, and Zheng et al.~\cite{zheng2019helen} propose Helen, a system that uses HE~\cite{paillier1999public} and verifiable secret sharing~\cite{damgaard2012multiparty} to execute ADMM~\cite{boyd2011distributed} (alternating direction method of multipliers, a convex optimization approach for distributed data), which supports regularized linear models. Similarly, Froelicher et al.~\cite{Drynx} employ HE~\cite{elgamal1985public}, along with encoding techniques, to enable the training of basic regression models and provide auditability with the use of zero-knowledge proofs. \sys enables better scalability in terms of the number of model's features, size of the dataset and number of data providers, and it offers richer functionalities by relying on the generic and widely-applicable SGD.

Another line of research considers the use of differential privacy for training ML models. Early works ~\cite{abadi2016deep, chaudhuri2009privacy} focus on a centralized setting where a trusted party holds the data, trains the ML model, and performs the noise addition. Differential privacy has also been envisioned in distributed settings, where to collectively train a model, multiple parties exchange or send differentially private model parameters to a central server ~\cite{du2018privacy, huang2019dp, Nvidia_Fed, shokri2015privacy}. However, the training of an accurate collective model requires very high privacy budgets and, as such, it is unclear what privacy protection is achieved in practice~\cite{jayaraman2019evaluating, Wang2019,hitaj2017deep}. To this end, some works consider hybrid approaches where differential privacy is combined with HE~\cite{kim2020,pathak2010multiparty}, or multi-party computation techniques~\cite{jayaraman2018distributed,truex2019hybrid}. \dfm{We consider differential privacy as an orthogonal approach; these techniques can be combined with our solution to protect the resulting models and their predictions from inference attacks~\cite{fredrikson2015model,shokri2017membership}, see Section \ref{extension_malicious}.}

\descr{Privacy-Preserving Prediction on ML Models.} Another line of work is focused on privacy-preserving ML prediction, where a party (e.g., a cloud provider) holds an already trained ML model on which another party (e.g., a client) wants to evaluate its private input. In this setting, Bost et al.~\cite{bost2015machine} use additive HE techniques to evaluate naive Bayes and decision tree classifiers, whereas Gilad-Bachrach et al.~\cite{gilad2016cryptonets} employ fully homomorphic encryption (FHE)~\cite{bos2013improved} to perform prediction on a small neural network. The computation overhead of these approaches has been further optimized by using multi-party computation (MPC) techniques~\cite{riazi2018chameleon,rouhani2018deepsecure}, or by combining HE and MPC~\cite{juvekar2018gazelle, liu2017oblivious,delphi}. Riazi et al.~\cite{riazi2019xonn} evaluate deep neural networks by employing garbled circuits and oblivious transfer, in combination with binary neural networks. Boemer et al.~\cite{boemer2019ngraph} propose nGraph-HE2, a compiler that enables service providers to deploy their trained ML models in a privacy-preserving manner. Their method uses HE, or a hybrid scheme that combines HE with MPC, to compile ML models that are trained with well-known frameworks such as TensorFlow~\cite{tensorflow} and PyTorch~\cite{pytorch}. The scope of our work is broader than these approaches, as \sys accounts not only for the private evaluation of machine-learning models but also for their privacy-preserving training in the distributed setting.

%% file: system.tex
\section{Secure Federated Training and Evaluation}\label{sec:overview}

We first introduce the problem of privacy-preserving distributed training and evaluation of machine-learning (ML) models. Then, we present a high-level overview and architecture of a solution that satisfies the security requirements of the presented problem. In Section~\ref{systemDesign}, we present \df{\sys, a system that enables the privacy preserving and distributed execution of a stochastic gradient-descent. We instantiate our solution for }the training and evaluation of the widely-used Generalized Linear Models~\cite{GLM}. In the rest of this paper, matrices are denoted by upper-case-bold characters and vectors by lowercase-bold characters; the i-th row of a matrix $\bm{X}$ is depicted as $\bm{X}[i,\cdot]$, and its i-th column as $\bm{X}[\cdot,i]$. Similarly, the i-th element of a vector $\bm{y}$ is denoted by $\bm{y}[i]$. We provide a list of recurrent symbols in Table~\ref{TableNotations} (see Appendix~\ref{appendixNotation}).

\subsection{Problem Statement}\label{sec:prob_form}

We consider a setting where a dataset $(\bm{X}_{n \times c}, \bm{y}_n)$, with $\bm{X}_{n \times c}$ a matrix of $n$ records and $c$ features, and $\bm{y}_n$ a vector of $n$ labels, is distributed among a set of data providers, i.e., $S =\{DP_1, \dots, DP_{|S|} \}$. The dataset is horizontally partitioned, i.e., each data provider $DP_i$ holds a partition of $n_i$ samples $(\bm{X}^{(i)}, \bm{y}^{(i)})$, with $\sum_{i=1}^{|S|} n_i = n$. A querier, which can also be a data provider (DP), requests the training of a ML model on the distributed dataset $(\bm{X}_{n \times c}, \bm{y}_n)$ or the evaluation of an already trained model on its input $(\bm{X'}, \cdot)$.

\begin{figure}[t]
\vspace{-1.5em}
    \centering
    \small
    \includegraphics[width=0.45\columnwidth]{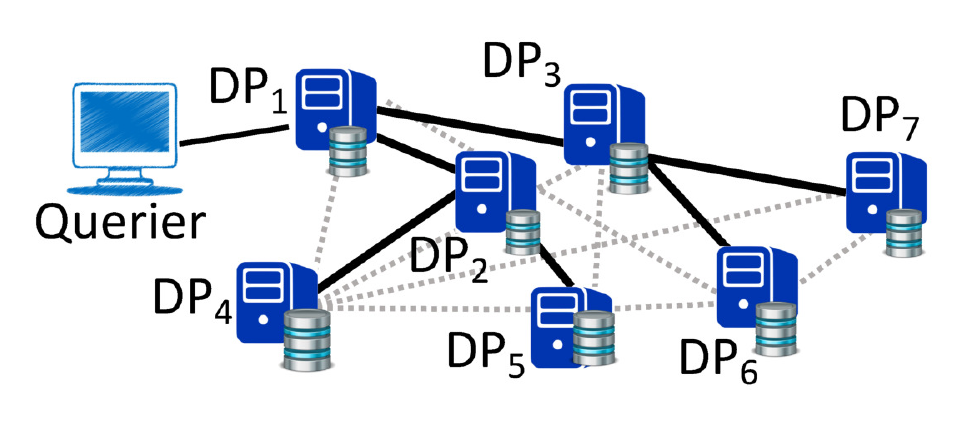}
    \vspace{-1.5em}
    \caption{\small{\sys's Model. Thick arrows represent a possible (efficient) query-execution flow.}}
    \label{fig:systemModel}

\end{figure}

We assume that the DPs are willing to contribute their respective data to train and to evaluate ML models on the distributed dataset. To this end, DPs are all interconnected and organized in a topology that enables efficient \df{execution of the computations, e.g., in a tree structure as depicted in Figure~\ref{fig:systemModel}}. 
Even though the DPs wish to collaborate for the execution of ML workflows, they do not trust each other. As a result, they seek to protect the confidentiality of their data (used for training and evaluation) and of the collectively learned model. \df{More formally, we require that the following privacy properties hold in a passive-adversary model in which all-but-one DPs can collude, i.e., the DPs follow the protocol, but up to $|S|-1$ DPs might share among them the information they observe during the execution, to extract information about the other DPs' inputs.}

\descr{(a) Data Confidentiality:} The training data of each data provider $DP_i$, i.e., $(\bm{X}^{(i)}, \bm{y}^{(i)})$ and the querier's  evaluation data $(\bm{X'}, \cdot)$ should remain only known to their respective owners. To this end, data confidentiality is satisfied as long as the involved parties (DPs and querier) do not obtain any information about other parties' inputs other than what can be deduced from the output of the process of training or evaluating a model.

\descr{(b) Model Confidentiality:} During the training process, no data provider $DP_i$ should gain more information about the model that is being trained than what it can learn from its own input data $(\bm{X}^{(i)}, \bm{y}^{(i)})$. During prediction, the querier should not learn anything more about the model than what it can infer from its input data $(\bm{X'}, \cdot)$ and the corresponding predictions $\bm{y'}$.

\dfm{\df{We remark here that input correctness and computation correctness are not part of the problem requirements, i.e., we assume that DPs input correct data and do not perform wrong computations. 
We discuss possible countermeasures against malicious DPs in Section~\ref{extension_malicious}.}}

\subsection{Solution Overview}\label{sec:sys_overview}

To address the problem of privacy-preserving distributed learning, we leverage the MapReduce abstraction, which is often used to capture the parallel and repetitive nature of distributed learning tasks~\cite{chu2007map,verbraeken2019survey}. \df{We complement this abstraction with a protection mechanism $P(\cdot)$; $P(x)$ denotes that value $x$ has to be protected to satisfy data and model confidentiality (Section~\ref{sec:prob_form}). We present the extended MapReduce abstraction in Protocol~\ref{alg:mapReduce}. In \textsc{prepare}, the data providers ($DP_i \in S$) pre-process their data $(\bm{X}^{(i)},\bm{y}^{(i)})$, \dfm{they agree on the learning parameters and on one data provider that plays the role of $DP_R$ and is then responsible for the execution of \textsc{reduce}. As explained later, $DP_R$ only manipulates protected data and is subject to the same security constraints as any other DP. We discuss the choice of $DP_R$ and its availability in Section \ref{sec:extensions}}. Each $DP_i$ then iteratively ($g$ iterations) trains its local model ($P(\bm{W}^{(i, j)})$ at iteration $j$) on its data in \textsc{map}. They combine their local models in \textsc{combine} (through an application-dependent function $C(\cdot)$), and update the global model $P(\bm{W}_G^{(\cdot, j)})$ in \textsc{reduce}. To capture the complete ML workflow, we extend the MapReduce architecture with a \textsc{prediction} phase in which predictions $P(\bm{y'})$ are computed from the querier's protected evaluation data $P(\bm{X'})$ by using the (protected) global model $P({\bm{W}_{G}}^{(\cdot, g)})$ obtained during the training.}
\begin{figure}[h]
\begin{protocol}[H]
    \small
	\caption{Extended MapReduce Abstraction.}
	\renewcommand{\thealgorithm}{}
	\begin{algorithmic}[1]
	\vspace{-0.5em} 
    \item[\textsc{training}: $S$ receives query from Querier and outputs $P(\bm{W}_G^{(\cdot,g)})$]
	\STATE Each $DP_i$ has $(\bm{X}^{(i)}$, $\bm{y}^{(i)})$ 
	\STATE \dfm{DPs appoint $DP_{R}$ and agree on learning params.} \hfill -- \textsc{prepare}
	\STATE Each $DP_i \in S$ initializes its local model $\bm{W}^{(i,0)}$
	\sbline
	\STATE{\textbf{for} $j = 1, \dots, g$ \textbf{do}}
	\STATE\hspace{0.3em} Each $DP_i \in S$ computes:  \hfill -- \textsc{map} \\
	\quad $P({\bm{W}}^{(i,j)}) \leftarrow \text{Map}((\bm{X}^{(i)},\bm{y}^{(i)}), P(\bm{W}_G^{(\cdot,j-1)}), P(\bm{W}^{(i, j-1)}))$
	\sbline
	\STATE{\hspace{0.3em} Each $DP_i$ sends $P(\bm{W}^{(i,j)})$ to $DP_{R}$ \hfill -- \textsc{combine}}
	\STATE{\hspace{0.32em}$DP_R$ computes: $P(\bm{W}^{(\cdot,j)})$ $\leftarrow$ $C(P(\bm{W}^{(i,j)})),$ $\forall$ $DP_i$ $\in$ $S$}
	\item[$ $ \hspace{52em} -- \textsc{reduce}]
	\STATE{\hspace{0.32em}$DP_R$ computes: $P({\bm{W}_G}^{(\cdot, j)})$$\leftarrow$$\text{Red}(P({\bm{W}_G}^{(\cdot, j-1)}), P(\bm{W}^{(\cdot,j)}))$}
	\sbline 
	\hspace{-0.2cm}{\textsc{prediction}: \dfm{$DP_R$ receives $P(\bm{X'})$ from Querier and uses $P({\bm{W}_G}^{(\cdot, g)})$} to compute $P(\bm{y'})$ that is sent back to the Querier}
	\vspace{-0.3em}
	\end{algorithmic}
	\label{alg:mapReduce}
\end{protocol}
\end{figure}

\section{SPINDLE Design}\label{systemDesign}

Following the extended MapReduce abstraction described in Section~\ref{sec:sys_overview}, we design a system, named \sys, \df{that enables the privacy-preserving execution of the widely applicable cooperative gradient descent~\cite{Wang_CooperativeSGD,Wang_CooperativeSGD_pres} -- which is used to minimize many cost functions in machine-learning~\cite{Kumar_2015,toulis2014statistical,Zhang_2004}}. We instantiate this system for the training of and prediction on Generalized Linear Models~\cite{GLM}. To implement the protection mechanism $P(\cdot)$, it builds on a multiparty fully homomorphic encryption scheme. We introduce these concepts in Section~\ref{sec:buildingBlocks}. Then, in Section~\ref{subsec:sys_proto}, we describe how \sys instantiates the phases of the extended MapReduce abstraction and how we address the collective data-processing on the distributed dataset through secure and interactive protocols. We also demonstrate how training is performed, notably by executing the gradient descent operations under homomorphic encryption, and how predictions are executed on encrypted models. The detailed cryptographic operations are presented in Section~\ref{sec:optimizations}.

\subsection{Background}\label{sec:buildingBlocks}

\df{\descr{Cooperative Gradient-Descent.}} We rely on a distributed version of the popular mini-batch stochastic gradient-descent (SGD)~\cite{Kumar_2015,toulis2014statistical,Zhang_2004}. In the standard version of SGD, the goal is to minimize $\min_{\bm{w}} [F(\bm{w}) := (1/n) \textstyle{\sum_{\phi=1}^{n}}\ f(\bm{w};\ \bm{X}[\phi,\cdot])]$, where $f(\cdot)$ is the loss function defined by the learning model, $\bm{w}$ $\in$ $\mathbb{R}^c$ are the model parameters, and $\bm{X}[\phi,\cdot]$ is the $\phi^{th}$ data sample (row) of $\bm{X}$. The model is then updated by $m$ iterations $\bm{w}^{(l)} = \bm{w}^{(l-1)} - \alpha [\zeta(\bm{w}^{(l-1)};\ \bm{B}^{(l)})],$ for $l$ $=$ $1$, $\dots$, $m$, with $\alpha$ the learning rate, $\bm{B}^{(l)}$ a randomly sampled sub-matrix of $\bm{X}$ of size $b \times c$, and $\zeta(\bm{w};\bm{B}) = \bm{B}^{T}(\sigma(\bm{B}\bm{w}) - I(\bm{z}))$, where $\bm{z}$ is the vector of labels corresponding to the batch $\bm{B}$. \df{The activation function $\sigma$ and $I(\cdot)$ are both model-dependent, e.g., for a logistic regression $\sigma$ is the sigmoid and $I(\cdot)$ is the identity.} 

We rely on the \textbf{cooperative SGD} (CSGD) proposed by Wang and Joshi~\cite{Wang_CooperativeSGD,Wang_CooperativeSGD_pres}, due to its properties; in particular: (i) modularity, as it can be synchronous or asynchronous, and can be combined with classic gradient-descent convergence optimizations such as Nesterov accelerated SGD \cite{nesterov2005smooth}; (ii) applicability, as it accommodates any ML model that can be trained with SGD and enables the distribution of any SGD based solution; (iii) it guarantees a bound on the error-convergence depending on the distributed parameters; e.g., the number of iterations and the update function for the global weights~\cite{Wang_CooperativeSGD,Wang_CooperativeSGD_pres,bottou2018optimization,zhang2015deep}; and (iv) it has been shown to work well even in the case of non-independent-and-identically-distributed (non-i.i.d.) data partitions~\cite{mcmahan2016communication,Wang_CooperativeSGD,Wang_CooperativeSGD_pres}. The data providers (DPs), each of which owns a part of the dataset, locally perform multiple iterations of the SGD before aggregating their model weights into the global model weights. The global weights are included in subsequent local DP computations to avoid that they learn, or descend, in the wrong direction. For simplicity, we present \sys with the synchronous CSGD version, where the DPs perform local model updates simultaneously. For each $DP_i$, the local update rule at global iteration $j$ and local iteration $l$ is:
\begin{equation}
\hspace{-0.4em}\bm{w}^{(i,j,l)} \text{=} \bm{w}^{(i,j,l\text{-}1)} - \alpha \zeta(\bm{w}^{(i,j,l\text{-}1)};\bm{B}^{(l)}) - \alpha \rho (\bm{w}^{(i,j,l\text{-}1)}-\bm{w}_G^{(\cdot,j\text{-}1)}),
\label{equation_gradient}
\end{equation}
where $\bm{w}_G^{(\cdot,j-1)}$ are the global weights from the last global update iteration $j-1$, $\alpha$ is the learning rate and $\rho$, the elastic rate, is the parameter that controls how much the data providers can diverge from the global model. The set of DPs $S$ performs $m$ local iterations between each update of the global model that is updated at global iteration $j$ with a moving average by:
   \begin{equation}
    \bm{w}_G^{(\cdot,j)}  = (1-{|S|}\alpha\rho) \bm{w}_G^{(\cdot,j-1)} + \alpha\rho \textstyle{\sum_{i=0}^{|S|}}\bm{w}^{(i,j,m)}.  
    \label{equationSGD}
 \end{equation}

\df{\descr{Generalized Linear Models (GLMs).}} GLMs~\cite{GLM} are a generalization of linear models where the linear predictor, i.e., the combination $\bm{X}\bm{w}$ of the feature matrix $\bm{X}$ and weights vector $\bm{w}$, is related to a vector of class labels $\bm{y}$ by an activation function $\sigma$ such that $E(\bm{y}) = \sigma^{-1}(\bm{X}\bm{w}) $, where $E(\bm{y})$ is the mean of $\bm{y}$. In this work, we consider the widely-used linear (i.e., $\sigma(\bm{Xw}) = \bm{Xw}$), logistic (i.e., $\sigma(\bm{Xw}) = {1}/{(1+e^{-\bm{Xw}})}$) and multinomial (i.e., $\sigma(\bm{Xw}_{\lambda}) = {e^{\bm{Xw}_{\lambda}}}/{(\sum_{j \in cl} e^{\bm{Xw}_j})}$, for $\lambda \in cl$) regression models. We remark that for multinomial regression, the weights are represented as a matrix $\bm{W}_{c \times |cl|}$, where $c$ is the number of features, $cl$ is the set of class labels and $|cl|$ its cardinality. In the rest of the paper, unless otherwise stated, we define the operations on a single vector of weights $\bm{w}$ and we note that in the case of multinomial regression, they are replicated on the $|cl|$ vectors of weights, i.e., each column of $\bm{W}_{c \times |cl|}$.

\descr{\DHE .} For the protection mechanism of \sys, we rely on a multiparty (or distributed) fully-homomorphic encryption scheme~\cite{mouchet2019distributedbfv} in which the secret key is distributed among the parties, while the corresponding collective public key $pk$ is known to all of them. Thus, each party can independently compute on ciphertexts encrypted under $pk$ but all parties have to collaborate to decrypt a ciphertext. In \sys, this enables the data providers (DPs) to train a collectively encrypted model, that cannot be decrypted as long as one DP is honest and refuses to participate in the decryption. As we show later, this multiparty scheme also enables DPs to collectively switch the encryption key of a ciphertext from $pk$ to another public key without decrypting. In \sys, a collectively encrypted prediction result can thus be switched to the querier's public key, so that only the querier can decrypt the result.

\df{Mouchet et al.~\cite{mouchet2019distributedbfv} propose a multiparty version of the Brakerski Fan-Vercauteren (\textsc{bfv}) lattice-based homomorphic cryptosystem~\cite{fan2012somewhat} and introduce interactive (distributed) protocols for key generation $\text{DKeyGen}(\cdot)$, decryption $\text{DDec}(\cdot)$, and bootstrapping $\text{DBootstrap}(\cdot)$. We use an adaptation of this multiparty scheme to the Cheon-Kim-Kim-Song cryptosystem (\textsc{ckks})~\cite{cheon2017homomorphic} that enables approximate arithmetic, and whose security is based on the ring learning with errors (\textsc{rlwe}) problem~\cite{lyubashevsky2010ideal}.} 
\textsc{ckks} (See Appendix~\ref{app:mhe}) enables arithmetic over $\mathbb{C}^{N/2}$; the plaintext and ciphertext spaces share the same domain $R_Q=\mathbb{Z}_Q[X]/(X^N+1)$, with $N$ a power of 2. Both plaintexts and ciphertexts are represented by polynomials of $N$ coefficients (degree $N-1$) in this domain. A plaintext/ciphertext encodes a vector of up to $N/2$ values.

\descrit{Parameters:} The \textsc{ckks} parameters are denoted by the tuple $(N,\Delta,\eta, mc)$, where $N$ is the ring dimension, \dfm{$\Delta$ is the plaintext scale, or precision, by which any value is multiplied before being quantized and encrypted/encoded}, $\eta$ is the standard deviation of the noise distribution, and $mc$ represents a chain of moduli $\{q_0, \dots, q_L \}$ such that $\Pi_{\iota \in \{0, \dots, \tau \}} q_\iota = Q_\tau$ is the ciphertext modulus at level $\tau$, with $Q_L=Q$, the modulus of fresh ciphertexts. Operations on a level-$\tau$ ciphertext $\langle\bm{v}\rangle$ are performed modulo $Q_\tau$, with $\Delta$ always lower than the current $Q_\tau$. Ciphertexts at level $\tau$ are simply vectors of polynomials in $R_{Q_\tau}$, that we represent as $\langle\bm{v}\rangle$ when there is no ambiguity about their level, and use $\{\langle\bm{v}\rangle, \tau, \Delta\}$ otherwise. After performing operations that increase the noise and the plaintext scale, $\{\langle\bm{v}\rangle, \tau, \Delta\}$ has to be rescaled \df{(see the $\text{ReScale}(\cdot)$ procedure defined in Appendix \ref{app:mhe}}) and the next operations are performed modulo $Q_{\tau-1}$. When reaching level $0$, $\langle\bm{v}\rangle$ has to be bootstrapped. The security of the cryptosystem depends on the choice of $N$, $Q$ and $\eta$, which in this work are parameterized to achieve at least 128-bits of security.

\descrit{(Distributed) Operations:} \df{A vector $\bm{v}$ of cleartext values can be encrypted with the public collective key $pk$ and can be decrypted with the collaboration of all DPs ($\text{DDec}(\cdot)$ protocol, in which each $DP_i$ uses its secret key $sk_i$). The DPs can also change a ciphertext encryption from the public key $pk$ to another public key $pk'$ without decrypting the ciphertext, by relying on the $\text{DKeySwitch}(\cdot)$ protocol. Each DP can independently add, multiply, rotate (i.e., inner-rotation of $\bm{v}$), rescale $\text{Rescale}(\cdot)$ or relinearize $\text{Relin}(\cdot)$ a vector encrypted with $pk$.  When two ciphertexts are multiplied together, the result has to be relinearized $\text{Relin}(\cdot)$ to preserve the ciphertext size. After multiple $\text{Rescale}(\cdot)$ operations, $\langle\bm{v}\rangle$ has to be refreshed by a collective protocol, i.e., $\text{DBootstrap}(\cdot)$, which returns a ciphertext at level $L$. The dot product $\text{DM}(\cdot)$ of two encrypted vectors of size $a$ can be executed by a multiplication followed by $log_2(a)$ inner-left rotations and additions. We list all the operations used in \sys and their properties in Appendix \ref{app:mhe}.}

\subsection{SPINDLE Protocols}\label{subsec:sys_proto}

We first describe \sys's operations for training a Generalized Linear Model following Protocol~\ref{alg:mapReduce}. In this case, the model $\bm{W}$ is a vector of weights that we denote by $\bm{w}$, and \textsc{map} corresponds to multiple local iterations of the gradient descent. Recall that in the case of multinomial regression, all operations are repeated for each label class $\lambda$ $\in$ $cl$.

\subsubsection{TRAINING}\label{training}

\descr{PREPARE.} The data providers (DPs) collectively agree on the training parameters: the maximum number of global $g$ and local $m$ iterations, and the learning parameters $lp=\{\alpha, \rho, b\}$, where $\alpha$ is the learning rate, $\rho$ the elastic rate, and $b$ the batch size. The DPs also collectively initialize the cryptographic keys for the distributed \textsc{ckks} scheme by executing $\text{DKeyGen}(\cdot)$ (see Appendix~\ref{app:mhe}). 
Then, the DPs initialize their local weights and pre-compute operations that involve only their input data ($\alpha\bm{X}^{(i)}I(\bm{y}^{(i)})$ and $\alpha \bm{X}^{(i)T}$). We discuss in Section~\ref{sec:extensions} how the DPs can collaborate to standardize or normalize the distributed dataset (if needed) and check that their respective inputs are consistent, e.g., they have data distribution homogeneity.

\descr{MAP.} As depicted in Protocol~\ref{alg:training}, the DPs execute $m$ iterations of the cooperative gradient-descent local update (Section~\ref{sec:buildingBlocks}). The local weights of $DP_i$ (i.e., $\langle\bm{w}^{(i, j, l-1)}\rangle$) are updated at a global iteration $j$ and a local iteration $l$ by computing the gradient (Protocol~\ref{alg:training}, lines 4, 5, and 6) that is then combined with the current global weights $\langle\bm{w}_G^{(\cdot, j-1)}\rangle$ (Protocol~\ref{alg:training}, line 7) following Equation~\ref{equation_gradient}. These computations are performed on batches of $b$ samples and $c$ features. To ensure that the update of $DP_i$'s local weights, i.e., the link between the ciphertexts $\langle\bm{w}^{(i,j-1)}\rangle=\langle\bm{w}^{(i,j,0)}\rangle$ and $\langle\bm{w}^{(i,j,m)}\rangle$, does not leak information about the DP's local data, $\langle\bm{w}^{(i,j,m)}\rangle$ is re-randomized $\text{RR}(\cdot)$ at the end of \textsc{map}, i.e., $DP_i$ adds to it a fresh encryption of 0. 
\begin{figure}[h]
\vspace{-1.0em}
\begin{protocol}[H]
    \small
	\caption{\textsc{map}.}
	\label{alg:training}
	\renewcommand{\thealgorithm}{}
	\begin{algorithmic}[1]
	\item[Each $DP_i$ outputs $\langle{\bm{w}}^{(i,j)}\rangle$$\leftarrow$$\text{Map}((\bm{X}^{(i)}, \bm{y}^{(i)}), \langle\bm{w}_G^{(\cdot,j-1)}\rangle,\langle\bm{w}^{(i, j-1)}\rangle)$]
	\STATE $\langle\bm{w}^{(i,j,0)}\rangle = \langle\bm{w}^{(i, j-1)}\rangle$
	\STATE \textbf{for} {$l = 1, \dots, m$} \textbf{:}
	\STATE \quad\text{Select batch ($\bm{B},\bm{z}$) of $b$ rows in $(\bm{X}^{(i)}, \bm{y}^{(i)})$}
		\STATE \quad $\langle\bm{u}[k]\rangle = \text{DM}(\bm{B}[k, \cdot], \langle{\bm{w}}^{(i,j,l-1)}\rangle)$, for $k$ $=$ $1$, $\dots$, $b$
		\sblinesmall
		\STATE \quad $\langle\bm{v}[e]\rangle = \text{DM}(\alpha \bm{B}[\cdot, e]^T, \sigma(\langle\bm{u}\rangle))$, \,\,for $e=1,\dots, c$
		\sblinesmall
		\STATE \quad $\bm{\mu}[e] = \sum_{k = 1}^{b} \alpha \bm{B}[\cdot, e]^T I(\bm{z}[k])$, \,\,\,\,\,\,for $e = 1, \dots, c$
		\sblinesmall
		\STATE \quad $\langle\bm{w}^{(i,j,l)}\rangle$ $=$ $\langle\bm{w}^{(i,j,l-1)}\rangle$+$\bm{\mu}$-$\langle\bm{v}\rangle$-$\alpha\rho(\langle\bm{w}^{(i,j,l-1)}\rangle$-$\langle{\bm{w}_G}^{(\cdot, j-1)}\rangle)$
	\STATE $\langle \bm{w}^{(i, j)} \rangle = \text{RR}(\langle \bm{w}^{(i, j, m)} \rangle$)
	\vspace{-0.3em}
	\end{algorithmic}
\end{protocol}
\vspace{-1.0em}
\end{figure}

Note that in Protocol~\ref{alg:training}, line 5 the activation function $\sigma(\cdot)$ is computed on the encrypted vector $\langle\bm{u}\rangle$ (or a matrix $\langle\bm{U}\rangle$ in the case of multinomial). The exponential activation functions for logistic (i.e., sigmoid) and multinomial (i.e., softmax) regressions have to be approximated to polynomial functions to be evaluated on encrypted data by using the homomorphic properties of \textsc{ckks}. We rely on a least-square polynomial approximation (LSPA) for the sigmoid, as it provides an optimal average mean-square error for uniform inputs in a specific interval, which is a reasonable assumption when the input distribution is not known. For softmax, we rely on Chebyshev approximation (CA) to minimize the maximum approximation error and thus avoid that the function diverges on specific inputs.
The approximation intervals can be empirically determined by using synthetic datasets with distribution similar to the real ones, by computing the minimum and maximum input values over all DPs and features, or by relying on estimations based on the data distribution~\cite{hesamifard2018privacy}. 
\begin{figure}[h]
\vspace{-1.0em}
\begin{protocol}[H]
    \small
	\caption{Activation Function $\sigma(\cdot)$.}
	\label{alg:activation}
	\renewcommand{\thealgorithm}{}
	\begin{algorithmic}[1]
	\vspace{-0.3em}
	\item[Func. $\sigma(\langle\bm{u}\rangle$\,or\,$\langle\bm{U}\rangle, t)$ returns the activated  $\langle\sigma(\bm{u})\rangle$ or $\langle\sigma(\bm{U})\rangle$]
	    \STATE{\textbf{if} $t$ is Linear \textbf{then} $\langle\sigma(\bm{u})\rangle = \langle\bm{u}\rangle$}
        \STATE{\textbf{else if} $t$ is Logistic \textbf{then}}
	    \STATE \hspace{0.1em} $\langle\sigma(\bm{u})\rangle = \text{apSigmoid}(\bm{u})$
	    \sblinesmall
	    \STATE{\textbf{else if} $t$ is Multinomial, input is a matrix $\langle\bm{U}_{c \times |cl|}\rangle$ \textbf{then}}
	    \STATE \hspace{0.1em} $\langle\bm{m}\rangle = \text{apMax}(\langle\bm{U}\rangle)$
	    \STATE \hspace{0.1em}\hspace{0.1em} \textbf{for} $\lambda$ $\in$ $cl$\textbf{:}
	    \STATE{ \hspace{0.1em}\hspace{0.1em}\hspace{0.1em} $\langle\bm{U'}[\lambda,\cdot]\rangle = \langle\bm{U}[\lambda,\cdot]\rangle-\langle \bm{m}\rangle$}
	    \STATE \hspace{0.1em}\hspace{0.1em}\hspace{0.1em}
	    $\langle\sigma(\bm{U}[\lambda,\cdot])\rangle$$=$$\text{M}(\text{apSoftN}(\langle\bm{U'}[\lambda,\cdot]\rangle),\text{apSoftD}(\langle\bm{U'}[\lambda,\cdot]\rangle))$
	\end{algorithmic}
\end{protocol}
\captionsetup{labelformat=empty}
\vspace{-1.0em}
\end{figure}
Protocol~\ref{alg:activation} takes as input an encrypted vector/matrix $\langle\bm{u}\rangle$ or $\langle\bm{U}\rangle$ and the type of the regression $t$ (i.e., linear, logistic or multinomial). If $t$ is linear, the protocol simply returns $\langle\bm{u}\rangle$. Otherwise, if $t$ is logistic, it computes the activated vector $\langle\sigma(\bm{u})\rangle$ by using the sigmoid's LSPA ($\text{apSigmoid}(\cdot)$). If $t$ is multinomial, it computes the activated matrix $\langle\sigma(\bm{U})\rangle$ using the softmax approximation that is computed by the multiplication of two CAs, one for the nominator $e^x$ ($\text{apSoftN}(\cdot)$) and one for the denominator $\frac{1}{\sum e^{x_j}}$ ($\text{apSoftD}(\cdot)$), each computed on different intervals. The polynomial approximation computation is detailed in Protocol~\ref{alg:OptimizedActivation} \df{(Appendix \ref{app:OptimizedActivation})}. To avoid an explosion of the exponential values in the softmax, a vector $\langle\bm{m}\rangle$ that contains the approximated max ($\text{apMax}(\cdot)$) value of each column of $\langle\bm{U}\rangle$ is subtracted from all input values, i.e., from each $\langle\bm{U}[\lambda,:]\rangle$ with $\lambda = 0,...,|cl|$. Similar to softmax, the approximation of the max function requires two CAs, and is detailed in Appendix~\ref{app:OptimizedActivation}.

\descr{COMBINE.} The \textsc{map} outputs of each $DP_i$, i.e., $\langle\bm{w}^{(i,j)}\rangle$, are homomorphically combined ascending a tree structure, such that each $DP_i$ aggregates its encrypted updated local weights with those of its children and sends the result to its parent. In this case, the combination function $C(\cdot)$ is the homomorphic addition operation. At the end of this phase, the DP at the root of the tree $DP_R$ obtains the encrypted combined weights $\langle\bm{w}^{(\cdot, j)}\rangle$.

\descr{REDUCE.} $DP_R$ updates the encrypted global weights $\langle\bm{w}_G^{(\cdot, j)}\rangle$, as shown in Protocol~\ref{alg:reduce}. More precisely, it computes Equation~\ref{equationSGD} by using the encrypted sum of the DPs' updated local weights $\langle{\bm{w}}^{(\cdot,j)}\rangle$ (obtained from \textsc{combine}), the previous global weights $\langle\bm{w}_G^{(\cdot, j-1)}\rangle$, the pre-defined elastic rate $\rho$ and the learning rate $\alpha$. After $g$ iterations of the \textsc{map}, \textsc{combine}, and \textsc{reduce}, $DP_R$ obtains the encrypted global model $\langle\bm{w}_G^{(\cdot, g)}\rangle$ and broadcasts it to the rest of the DPs.
\begin{figure}[h!]
\vspace{-1.0em}
\begin{protocol}[H]
    \small
	\caption{\textsc{Reduce.}}
	\label{alg:reduce}
	\renewcommand{\thealgorithm}{}
	\begin{algorithmic}[1]
	\vspace{-0.3em}
	\item[$DP_R$ computes $\langle\bm{w}_G^{(\cdot,j)}\rangle \leftarrow \text{Red}(\langle\bm{w}_G^{(\cdot, j-1)}\rangle, \langle\bm{w}^{(\cdot, j)}\rangle, \rho, \alpha)$]
	\STATE $\langle\bm{w}_G^{(\cdot,j)}\rangle = (1 -  \alpha \rho |S|) \langle\bm{w}_G^{(\cdot, j-1)}\rangle + \alpha \rho \langle\bm{w}^{(\cdot,j)}\rangle$
	\vspace{-0.3em}
	\end{algorithmic}
\end{protocol}
\vspace{-1.0em}
\end{figure}

\subsubsection{PREDICTION}\label{subsec:weigthsdisclosure}

The querier's input data $(\bm{X'}, \cdot)$ is encrypted with the collective public key $pk$. Then, $\langle\bm{X'}\rangle_{pk}$ is multiplied ($DM(\cdot, \cdot)$ with the weights of the trained model $\langle\bm{w}_G^{(\cdot, g)}\rangle$ and processed through the activation function $\sigma(\cdot)$ to obtain the encrypted prediction values $\langle\bm{y'}\rangle$ (one prediction per row of $\bm{X'}$). The prediction results encrypted under $pk$ are then collectively switched by the DPs to the querier public key $pk'$ using $\text{DKeySwitch}(\cdot)$, so that only the querier can decrypt $\langle\bm{y'}_{pk'}\rangle$.
\begin{figure}[h!]

\begin{protocol}[H]
\vspace{-1.0em}
	\caption{\textsc{prediction}.}
	\label{alg:inference}
	\renewcommand{\thealgorithm}{}
	\begin{algorithmic}[1]
	\item[$DP_R$ gets $\langle \bm{X'}_{n' \times c} \rangle$ from Querier and computes $\langle{\bm{y'}_{n'}}\rangle$ using $\langle\bm{w}_G^{(\cdot, g)}\rangle$]
	\STATE { $\langle \bm{y'}[p]\rangle = \sigma(\text{DM}(\langle\bm{X'}[p,\cdot]\rangle, {\langle\bm{w}_G^{(\cdot,g)}}\rangle))$, for $p=0,...,n'$}
	\STATE $\langle \bm{y'}\rangle_{pk'} = \text{DKeySwitch}(\langle \bm{y'}\rangle, pk',\{sk_i\})$
	\vspace{-0.2em}
	\end{algorithmic}
\end{protocol}
\vspace{-1.0em}
\end{figure}

%% file: optimizations.tex
\section{System Operations}\label{sec:optimizations}

We describe how \sys relies on the properties of the distributed version of \textsc{ckks} to efficiently address the problem of privacy-preserving distributed learning. We first describe how we optimize the protocols of Section~\ref{subsec:sys_proto} by choosing when to execute cryptographic operations such as rescaling and (distributed) bootstrapping. Then, we discuss how to efficiently perform the \textsc{map} protocol that involves a sequence of vector-matrix-multiplications and the evaluation of the activation function, in the encrypted domain.
\begin{figure*}[t]
    \centering
    \small
    \includegraphics[width=1\columnwidth]{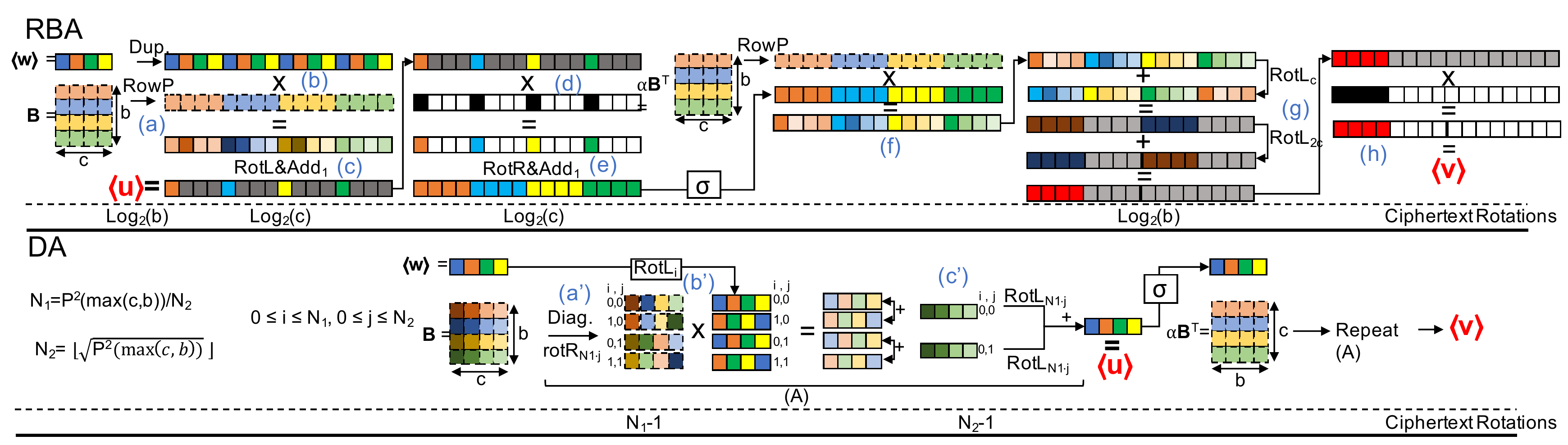}
    \vspace{-1em}
    \caption{\small{Packing approaches for executing Protocol~\ref{alg:training}, lines 4 and 5. We assume that $c \cdot b < N/2$ and show an example with $c=b=4$. Dash elements are plaintext values, everything else is encrypted. $\text{Dup}$ duplicates and adds, $\text{rowP}$ packs the rows in one ciphertext, $\text{RotL(/R)}\&\text{Add}_i$ rotates the encrypted vector by $i, 2i, 4i, \dots$ to the left(/right) and at each step, aggregates the result with the previous ciphertext, $\text{RotL(/R)}_j$ rotates a vector left(/right) by $j$ positions. $\text{P}^2(x)$ returns the next power of 2 larger than $x$.}}
    \label{fig:packingSchemes}
    \vspace{-2.0em}
\end{figure*}

\subsection{Cryptographic Operations}\label{sec:cryptoOperations}

\df{As explained in Section~\ref{sec:buildingBlocks} (and Appendix \ref{app:mhe}), ciphertext multiplications incur the execution of other cryptographic operations hence increase \sys's computation overhead.} This overhead can rapidly increase when the same ciphertext is involved in sequential operations, i.e., when the operations' multiplicative depth is high. As we will describe in Section~\ref{sec:eval}, \sys relies on the Lattigo~\cite{lattigo} lattice-based cryptographic library, where a ciphertext addition or multiplication requires a few ms, whereas $\text{Rescale}(\cdot)$, $\text{Relin}(\cdot)$, and $\text{DBootstrap}(\cdot)$, are $1$-order, $2$-orders, and $1.5$-orders of magnitude slower than the addition, respectively. These operations can be computationally heavy, hence their execution in the protocols should be optimized. Note that we avoid the use of the centralized traditional bootstrapping, as it would require a much more conservative parameterization for the same security level, resulting in higher computational overheads (see Section \ref{sec:eval}).

\descr{Lazy Rescaling.} To maintain the precision of the encrypted values and for efficiency we rescale a ciphertext $\{\langle\bm{v}\rangle,\tau,\Delta\}$ only when $\Delta$ is close to $q_\tau$. Hence, we perform a $\text{ReScale}(\cdot)$ only when this condition is met after a series of consecutive operations.

\descr{Relinearization.} Letting the ciphertext size increase after every multiplication would add to the subsequent operations an overhead that is higher than the relinearization. Hence, to maintain the ciphertext size and degree constant, a $\text{Relin}(\cdot)$ operation is performed after each ciphertext-ciphertext multiplication. \pl{We here note that a $\text{Relin}(\cdot)$ operation can be deferred if doing so incurs in lower computational complexity (e.g., if additions performed after the ciphertext-ciphertext multiplications reduce the number of ciphertexts to relinearize).}

\descr{Bootstrapping.} In the protocols of Section~\ref{subsec:sys_proto}, we observe that the data providers' local weights and the model global weights ($\langle\bm{w}
\rangle$ and $\langle\bm{w}_G\rangle$, resp.) are the only persistent ciphertexts over multiple computations and iterations. They are therefore the only ciphertexts that need to be bootstrapped, and \dfm{we consider three approaches for this}. With \textbf{Local bootstrap (LB)}, each data provider (DP) bootstraps (calling a $\text{DBootstrap}(\cdot)$ protocol) its local weights, every time they reach level $\tau_b$ during the \textsc{map} local iterations and before the \textsc{combine}. As a result, the global weights are always combined with fresh encryptions of the local weights and only need to be bootstrapped after multiple \textsc{reduce}. Indeed, \textsc{reduce} involves a multiplication by a constant hence a $\text{Rescale}(\cdot)$. With \textbf{Global bootstrap (GB)}, we use the interdependency between the local and global weights, and we bootstrap only the global weights and assign them directly to the local weights. The bootstrapping is performed on the global weights during \textsc{reduce}. Thus, we modify \textsc{training} so that \textsc{map} operates on the (bootstrapped) global weights, i.e., $\langle\bm{w}^{(i,j-1)}\rangle = \langle{\bm{w}_G}^{(\cdot,j-1)}\rangle$, for a $DP_i$ at global iteration $j$. By following this approach, the number of bootstrap operations is reduced, with respect to the local approach, because it is performed by only one DP and depends only on the number of global iterations. However, it modifies the learning method, and it offers less flexibility, as the number of local iterations in \textsc{map} is constrained by the number of ciphertext multiplications required in each iteration and by the available ciphertext levels. \dfm{With \textbf{Hybrid bootstrap (HB)}, both GB and LB approaches are combined to reduce the total number of bootstrapping operations. The global weights are bootstrapped at each global iteration (GB) and the DPs can still perform many local iterations by relying on the LB.} In our experiments (Section~\ref{subsec:eval}), we observed that the effect on the trained model's accuracy depends mainly on the data and that, in most cases, enabling DPs to perform more local iterations \dfm{(LB and HB)} between two global updates yields better accuracy. \dfm{Even though LB incurs at least $|S|$ more executions of the $\text{DBootstrap}(\cdot)$, the DPs execute them in parallel and thus amortize the overhead on \sys's execution time. However, if the training of a dataset requires frequent global updates, then GB (or HB) achieves a better trade-off, see Section \ref{subsec:eval}. Taking into account these cryptographic transformations and the strategy to optimize their use in \sys, we explain how to optimize the required number of ciphertext operations.}

\subsection{MAP Vector-Matrix Multiplications}\label{sec:MatrixMultSGD}

As described in Section~\ref{sec:buildingBlocks}, each \textsc{ckks} ciphertext encrypts (or packs) a vector of values, e.g., 8,192 elements if the ring dimension is $N$ $=$ $2^{14}$. This packing enables us to simultaneously perform operations on all the vector values, by using a Single-Instruction Multiple Data (SIMD) approach for parallelization. To execute computations among values stored in different slots of the same ciphertext, e.g., an inner sum, we rely on ciphertext rotations that have a computation cost similar to a relinearization ($\text{Relin}(\cdot)$). Recall that for the execution of stochastic gradient-descent, each local iteration in \textsc{map} involves two sequential multiplications between encrypted vectors and cleartext matrices (Protocol~\ref{alg:training}, lines 4 and 5). As a result, packing is useful for reducing the number of vector multiplications and rotations needed to perform these operations. To this end, \sys integrates two packing approaches and automatically selects the most appropriate approach at each DP during the training. We now describe these two approaches and how to choose between them, depending on the settings, i.e., the learning parameters, the number of features, and the DP computation capabilities. Figure~\ref{fig:packingSchemes} depicts \sys's packing approaches for a toy example of the computation of $\langle\bm{u}\rangle$ (Protocol~\ref{alg:training}, line 4) whose result is activated (i.e., $\sigma(\langle\bm{u}\rangle)$) before used in the computation of $\langle\bm{v}\rangle$ (Protocol~\ref{alg:training}, line 5), for a setting with $c=b=4$. For clarity, we assume that a vector of $c$ (number of features) or $b$ (batch size) elements can be encoded in one ciphertext (or plaintext), i.e., $\text{max}(c,b) \leq N/2$.

\descr{Row-Based Approach (RBA).} This approach was proposed by Kim et al.~\cite{kim2018logistic}. The input matrices ($\bm{B}$ and $\alpha\bm{B}^T$) are packed row-wise, and multiple rows are packed in one plaintext ($(a)$ in the upper part of Figure~\ref{fig:packingSchemes}), i.e., the number of plaintexts required to encode the input matrix is $\ceil{\frac{c \cdot b \cdot 2}{N}}$. Each plaintext is then multiplied with a ciphertext containing the replicated weights' vector $(b)$, such that the number of replicas is equal to the number of rows in $\bm{B}$. To obtain the results of the dot products between each weights' vector and row of $\bm{B}$, a partial inner sum is performed by adding the resulting ciphertext with rotated versions of itself $(c)$. The values in between the dot product results are eliminated (i.e., masked) through a multiplication with a binary vector $(d)$, and the dot product results are duplicated in the ciphertext $(e)$ such that it can be activated ($\sigma(\cdot)$) and used directly for the multiplication with $\alpha\bm{X^}T$ $(f)$. The result is then rotated and added to itself $(g)$ such that it can be masked $(h)$ to obtain $\langle\bm{v}\rangle$. As shown in Figure \ref{fig:packingSchemes}, the total number of vector multiplications is $\ceil{\frac{c \cdot b \cdot 2}{N}} \cdot 4$, whereas the number of ciphertext rotations is $\ceil{\frac{c \cdot b \cdot 2}{N}} \cdot 2 \cdot (log(b)+log(c))$. This approach has a multiplicative depth of $a_m+4$, where $a_m$ denotes the depth of the activation function $\sigma(\cdot)$.

\descr{Diagonal Approach (DA).} This approach was presented by Halevi and Shoup~\cite{Helib} as an optimized homomorphic vector-matrix-multiplication evaluation. It optimizes the number of ciphertext rotations by transforming the input plaintext matrix $\bm{B}$. In particular, $\bm{B}$ is diagonalized, and each line is rotated ($(a')$ in lower part of Figure~\ref{fig:packingSchemes}) so that they can be independently multiplied with the (rotated) weights' vector $(b')$. The resulting ciphertexts are aggregated and rotated to obtain $\langle\bm{u}\rangle$ $(c')$, and a similar approach is used to compute $\langle\bm{v}\rangle$ after the activation. As shown in Figure~\ref{fig:packingSchemes}, DA only executes $2 \cdot ((N_1 - 1) + (N_2 - 1))$ rotations on the encrypted vector, with $N_1 = P^2(\text{max}(c,b))/N_2$ and $N_2 = \lfloor \sqrt{P^2(\text{max}(c,b))} \rfloor$, where $P^2(x)$ returns the next power of 2 larger than $x$. This approach involves $N_1 \cdot N_2$ plaintext-ciphertext multiplications on independent ciphertexts and does not require any masking, which results in a multiplicative depth of $a_m+2$. Therefore, this approach consumes fewer levels than RBA.

In both approaches, the number of rotations and multiplications depends on the batch size $b$ and the number of features $c$. 
DA almost always requires more multiplications than RBA and uses more rotations after a certain $c$ (e.g., if $b=8$, the break-even happens at $c=64$). However, as DA is \textit{embarrassingly parallelizable} for both multiplications and rotations (with rotations being the most time-consuming operations), the computations can be amortized on multiple threads. Taking this into account, \sys automatically chooses, based on $c$, $b$, and the number of available threads, the best approach at each DP. We analyze these trade-offs in Section~\ref{sec:eval}.

\subsection{Optimized Activation Function}\label{sec:OptimizedActivation}

\df{As described in Section~\ref{subsec:sys_proto}, to enable their execution under FHE, we approximate the sigmoid ($\text{apSigmoid}(\cdot)$) and softmax ($\text{apMax}(\cdot)$, $\text{apSoftN}(\cdot)$, $\text{apSoftD}(\cdot)$) activation functions with least-squares and Chebyshev polynomial approximations (PA), respectively. 
We adapt the baby-step giant-step algorithm introduced by Han and Ki~\cite{Han_BetterBootstrap} to enable the minimum-complexity computation of degree-$d$ polynomials (multiplicative depth of $\ceil{log(d)}$ for $d \leq 7$, and with depth $\ceil{log(d)+1}$ otherwise). Protocol~\ref{alg:OptimizedActivation} in Appendix \ref{app:OptimizedActivation} inductively computes the (element-wise) exponentiation of the encrypted input vector before recursively computing the polynomial approximation.}

\section{System Configuration}\label{sec:systemConfiguration}

We discuss how to parameterize \sys by taking into account the interdependencies between the input data, and the learning and cryptographic parameters. We then discuss two modular functionalities of \sys, namely \textit{data outsourcing} and \textit{model release}.

\begin{figure}[h]
\small
	\centering
	\includegraphics[width=0.6\columnwidth]{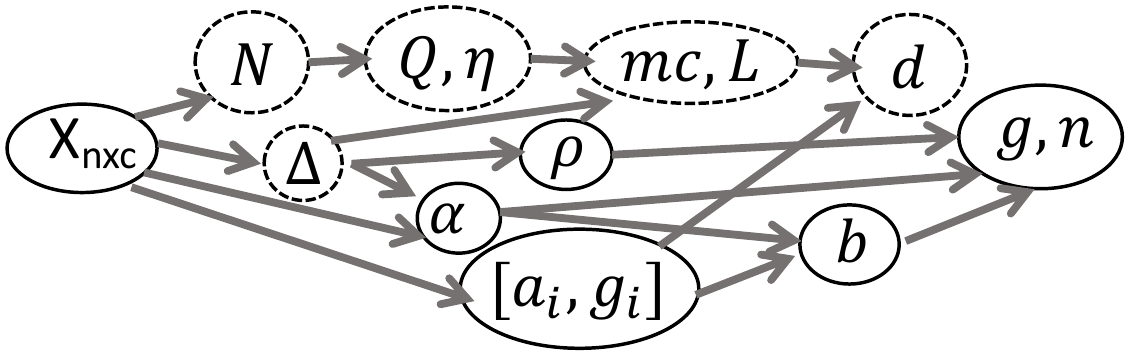}
	\vspace{-0.5em}
	\caption{\small{System parameters graph. Circles and dotted circles represent learning and cryptographic parameters, respectively.}}
	\label{fig:paramsWork}
\end{figure}

\descr{Parameter Selection.} \sys relies on the configuration of (a) cryptographic parameters that determine its security level, and (b) learning parameters that affect the accuracy of the training and evaluation of the models. Both are tightly linked, and we capture these relations in a graph-based model, displayed in Figure~\ref{fig:paramsWork}, where vertices and edges represent the parameters and their interdependence, respectively. For simplicity, we present a directed graph that depicts our empirical method for choosing the parameters (see Appendix~\ref{appendixNotation}, Table~\ref{TableNotations} for notation symbols). We highlight that the corresponding non-directed graph is more generic and simply captures the main relations among the parameters. We observe two main clusters: the cryptographic parameters on the upper part of the graph (dotted circles), and the learning parameters (circles) on the lower one. The input data and their intrinsic characteristics, i.e., the number of features $c$ or precision (bits of precision required to represent the data), are connected with both clusters that are also interconnected through the plaintext scale $\Delta$. As such, there are various ways to configure the overall system parameters.

In our case, we decide to first choose $N$ (ciphertext polynomial degree), such that at least $c$ elements can be packed in one ciphertext. $Q$ (ciphertext modulus) and $\eta$ (fresh encryption noise) are then fixed to ensure a sufficient level of security (e.g., 128-bits) following the accepted parameterization from the homomorphic encryption standard whitepaper~\cite{HEStandardPaper}. The scale $\Delta$ is configured to provide enough precision for the input data $\bm{X}$, and $mc$ (moduli chain) and $L$ (number of levels) are set accordingly. The intervals $[a_i,g_i]$ used for the approximations of the activation functions are defined according to $\bm{X}$. The approximation degrees $\bm{d}$ are then set depending on these intervals and the available number of levels $L$. The remaining learning parameters ($\alpha$, $\rho$, $b$, $g$, $m$) are agreed upon by the data providers based on their observation of their part of the dataset. Note that the minimum values for the learning rate $\alpha$ and elastic rate $\rho$ are limited by the scale $\Delta$, and if they are too small the system might not have enough precision to handle their multiplication with the input data.

\descr{Data Outsourcing.} \df{\sys's protocols (Section~\ref{subsec:sys_proto}) seamlessly work with data providers (DPs) that either have their input data $\bm{X}$ in cleartext, or that obtain data $\langle\bm{X}\rangle_{pk}$ encrypted under the public collective key from their respective owners. In the latter case, \sys enables both secure data storage and computation outsourcing to always-available untrusted cloud providers. It distributes the workload among multiple data providers and is still able to rely on efficient multiparty homomorphic-encryption operations, e.g., $\text{DBootstrap}(\cdot)$. We note that operating on encrypted input data affects the complexity of \textsc{map}, as all the multiplication operations (Protocol~\ref{alg:training}) would happen between ciphertexts, instead of between the cleartext inputs and ciphertexts.}



\descr{Model Release.} \df{By default, the trained model in \sys is kept secret from any entity, enabling privacy-preserving predictions on (private) evaluation-data input by the querier and offering end-to-end \textit{model confidentiality}. If required by the application setting, \sys can also reveal the trained model to the querier or to a third party. This is collectively enabled by the DPs, who perform a $\text{DKeySwitch}(\cdot)$.} 

%% file: security2.tex
\section{Security Analysis}
\label{sec:securityAnalysis}

\df{We demonstrate that \sys achieves the data and model confidentiality requirements defined in Section~\ref{sec:prob_form} by relying on the real/ideal simulation paradigm~\cite{lindell2017simulate} and showing that a computationally-bounded adversary that controls up to $(|S|-1)$-out-of-$|S|$ DPs cannot distinguish a \emph{real} world experiment, in which the adversary is given actual data (sent by honest DP(s)), and an \emph{ideal} world experiment, in which the adversary is given random data generated by a simulator.}

\df{The semantic security of the CKKS scheme is based on the hardness of the decisional RLWE problem~\cite{cheon2017homomorphic,lyubashevsky2010ideal,Lindner2011}. The achieved practical bit-security against state-of-the-art attacks can be computed using Albrecht's LWE-Estimator~\cite{HEStandardPaper,Albrecht2015OnTC}. Mouchet et al.~\cite{mouchet2019distributedbfv} prove that their distributed protocols, i.e., Collective Encryption-Key Generation, Collective Relinearization-Key Generation ($\text{DKeyGen}(\cdot)$) and Collective Key Switching ($\text{DKeySwitch}(\cdot)$ and $\text{DDec}(\cdot)$) are secure under the simulator paradigm. They show that the distribution of the cryptoscheme preserves its security in the passive adversary model with all-but-one dishonest DPs, as long as the decisional-\textsc{rlwe} problem is hard. Their proofs, which are constructed using the \textsc{bfv} scheme, generalize to our adaptation of their protocols to \textsc{ckks}, as they preserve the same computational assumptions, and the security of \textsc{ckks} is based on the same hard problem as \textsc{bfv}. The security of $\textsf{DBootstrap}(\cdot)$ is based on Lemma~\ref{lemmaDBoot} which we state and prove in Appendix~\ref{app:DBootstrapSecurity}.}
\begin{proposition}

\label{lemma1}
\df{Assume that \sys uses CKKS encryptions with parameters $(N,\Delta,\eta, mc)$ ensuring a post-quantum security level $\lambda$. Given a passive adversary corrupting at most $|S|-1$ parties, \sys achieves \emph{data and model confidentiality} for training and prediction.}
\end{proposition}
\descr{Sketch of the Proof.} \df{We consider a real-world simulator $\mathcal{S}$ that simulates the view of a computationally-bounded adversary corrupting $|S|-1$ parties, i.e., it has access to the inputs and outputs of $|S|-1$ parties. In \textsc{prepare} and \textsc{map}, the data providers (DPs) locally compute on their data and only exchange encrypted information with each other to perform $\text{DKeyGen}(\cdot)$ and $\text{DBootstrap}(\cdot)$. In \textsc{combine}, the DPs' \textsc{map} outputs, encrypted under the public collective key, are aggregated. These outputs are the encrypted results of multiple local iterations in which elements derived from each $DP_i$'s local private data $(\bm{X}^{(i)},\bm{y}^{(i)})$ are combined with its encrypted local model $\langle \bm{w}^{(i,\cdot)}\rangle$ and the current encrypted global model $\langle \bm{w}_G \rangle$. This result is re-randomized (i.e., added to a fresh encryption of $\bm{0}$) to ensure that the outputs of successive \textsc{map} (i.e., the inputs to \textsc{combine}) do not leak any information about the DPs' data. In \textsc{reduce}, the global model is updated by combining encrypted data, and a $\text{DBootstrap}(\cdot)$ is executed. All the information exchanged by the DPs is encrypted and the DPs rely on the aforementioned CPA-secure-proven protocols. We show in Appendix \ref{app:DBootstrapSecurity} that $\textsf{DBootstrap}(\cdot)$ is also simulatable. In \textsc{prediction}, only encrypted information is exchanged and the security-proven $\text{DKeyswitch}(\cdot)$ protocol is used. We consider two cases: (a) the adversary controls $|S|-1$ DPs and (b) it controls the querier and $|S|-2$ DPs. In (a), the encryption of the querier's input data (with the DPs common public key $pk$) can be simulated by $\mathcal{S}$ and \sys ensures \emph{Data Confidentiality} of the querier. In (b) the confidentiality of the adversary-controlled-querier's data is trivial. The simulator has access to the prediction result and can produce all the intermediate (indistinguishable) encryptions that the adversary sees. Hence, $\mathcal{S}$ can simulate all the values communicated during the \textsc{training} and \textsc{prediction} by generating random ciphertexts using the parameters $(N,\Delta,\eta, mc)$, such that the real outputs cannot be distinguished from the ideal ones. The sequential composition of all cryptographic functions remains simulatable by $\mathcal{S}$, as different random values are used in each step, and the exchanged ciphertexts are re-randomized, i.e., there is no dependency between the random values that an adversary can leverage on. Also, the adversary cannot decrypt collectively encrypted data unless all DPs collude, which would contradict the considered threat model. Following this, \sys ensures the data and model confidentiality of the honest party/ies.}

Finally, we note that by design, \sys thwarts active attacks on federated learning~\cite{hitaj2017deep, Melis2019, Nasr2019, NIPS2019_9617} and model inversion attacks~\cite{fredrikson2015model}, as intermediate and final model weights are never revealed during \textsc{training}.

%% file: evaluation.tex
\section{System Evaluation}\label{sec:eval}

We first analyze the theoretical complexity of \sys before moving to the empirical evaluation of its prototype and its comparison with existing solutions.

\subsection{Theoretical Analysis}

We refer to Table \ref{fig:theoryAnalysis} (Appendix \ref{app:evalcomplement}) for the full complexity analysis of \sys's protocols. We discuss here its main outcomes.

\descr{Communication Complexity.} \df{\sys's communication complexity depends linearly on the number of data providers $|S|$, iterations ($g,m$) and the ciphertext size $|ct|$. In \textsc{map}, the only communication between the DPs is due to the $\text{DBootstrap}(\cdot)$, which requires two rounds of communication of one ciphertext ($ct$) between the $|S|$ DPs (i.e., $2\cdot(|S|-1)\cdot|ct|$). In \textsc{combine} and \textsc{reduce}, the DPs exchange one ciphertext in respectively one and two rounds. Finally, the \textsc{prediction} requires the exchange of one ciphertext between a DP and the querier and one $\text{DKeySwitch}(\cdot)$ operation, i.e., 2 ciphertexts are sent per DP.}

\descr{Computation Complexity.} \df{\sys's most intensive computational part is \textsc{map}; its complexity depends linearly on the number of DPs $|S|$ and the number of iterations, and logarithmically on the number of features $c$ and batch size $b$; all these parameters depend also on the dataset size. As shown in Section \ref{sec:MatrixMultSGD}, the DA packing approach incurs a higher computation complexity but is embarrassingly parallel and can be more time-efficient than RBA depending on the available threads. \dfm{The activation function is the only operation that requires ciphertext-ciphertext multiplications; its complexity depends logarithmically on the approximation degree. We empirically study the link between the approximation degree and the training accuracy in Section \ref{subsec:eval}}. \sys's other steps and protocols only involve lightweight operations, i.e., ciphertexts additions and multiplications with plaintext values.}

\subsection{Empirical Evaluation}\label{subsec:eval}

We implemented \sys in Go~\cite{Go}. 
 Our implementation builds on top of Lattigo~\cite{lattigo}, an open-source Go library for lattice-based cryptography, and Onet~\cite{Onet}, an open-source Go library for building decentralized systems. The communication between data providers (DPs) is done through TCP with secure channels (using TLS). We evaluate our prototype on an emulated realistic network, with a bandwidth of \df{1~Gbps} between every two nodes, using Mininet~\cite{mininet}. We deploy \sys on 5 Linux machines with Intel Xeon E5-2680 v3 CPUs running at 2.5GHz with 24 threads on 12 cores and 256 Giga Bytes RAM, on which we evenly distribute the DPs.

\df{\dfm{We first provide \sys's cryptographic operations micro-benchmarks before assessing} \sys accuracy and performance by testing it on multiple publicly-available datasets: CalCOFI~\cite{CalCOFI} for linear regression, BCW~\cite{BCW}, PIMA~\cite{pima} and ESR~\cite{ESR} for logistic regression, and MNIST ~\cite{MNIST} for multinomial regression (see Appendix \ref{app:datasets} for details on the datasets). We then show \sys's scalability by using randomly generated (larger) datasets with up to 8,192 features and 4 million data samples. Our evaluation shows \sys practicality for large-dimensional datasets, making it suitable for demanding learning tasks such as the training on imaging or genomic datasets \cite{beam2018big,erickson2017machine,leung2015machine}}.

\begin{figure*}[h!]
	\centering
	\tiny
	\begin{subfigure}[t]{0.47\textwidth}
		\centering
		\includegraphics[width=1.0\columnwidth]
		{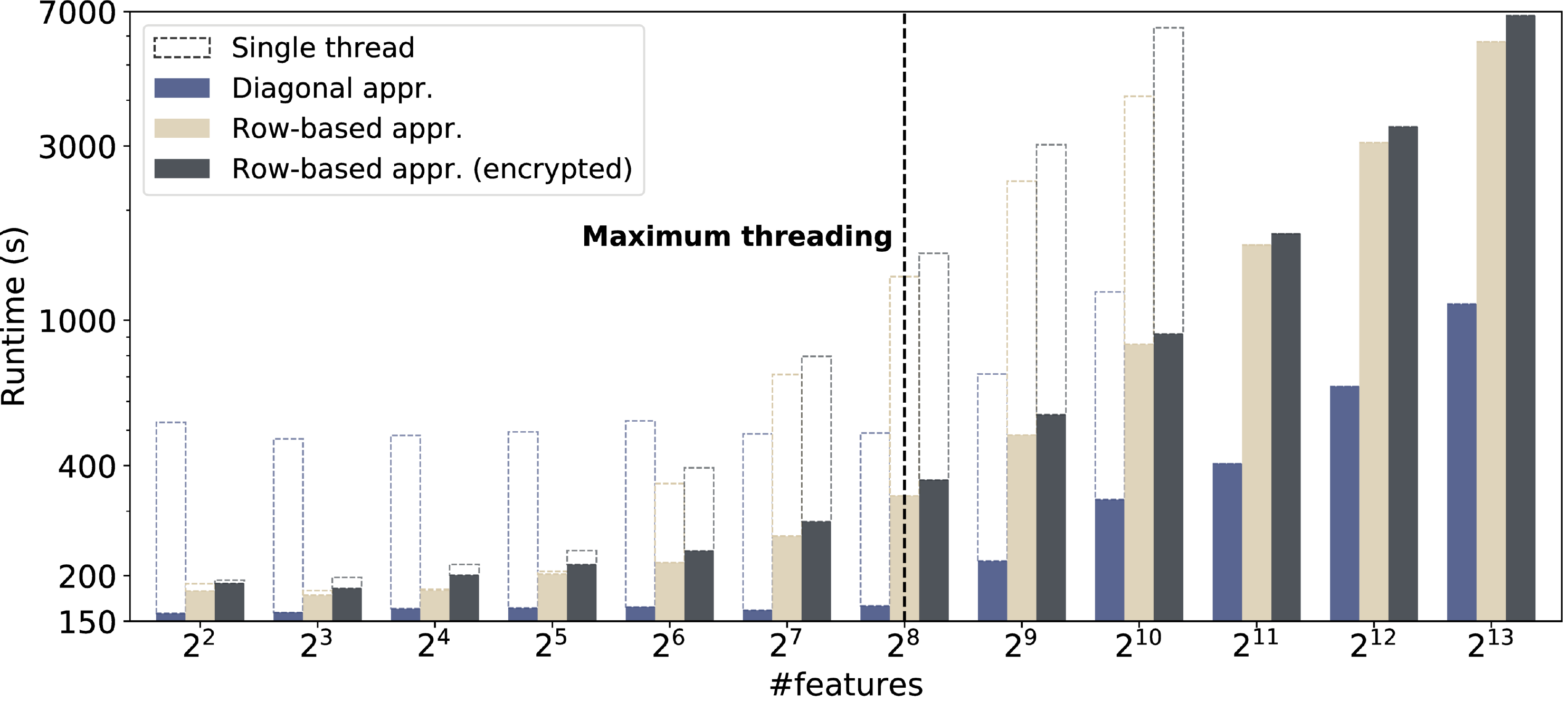}
		\caption{\sys's performance with the nbr. of features ($c$).}
		\label{fig:performanceFeature}
	\end{subfigure}
	\hspace{2em}
	\begin{subfigure}[t]{0.45\textwidth}
		\centering
		\includegraphics[width=1.0\columnwidth]
		{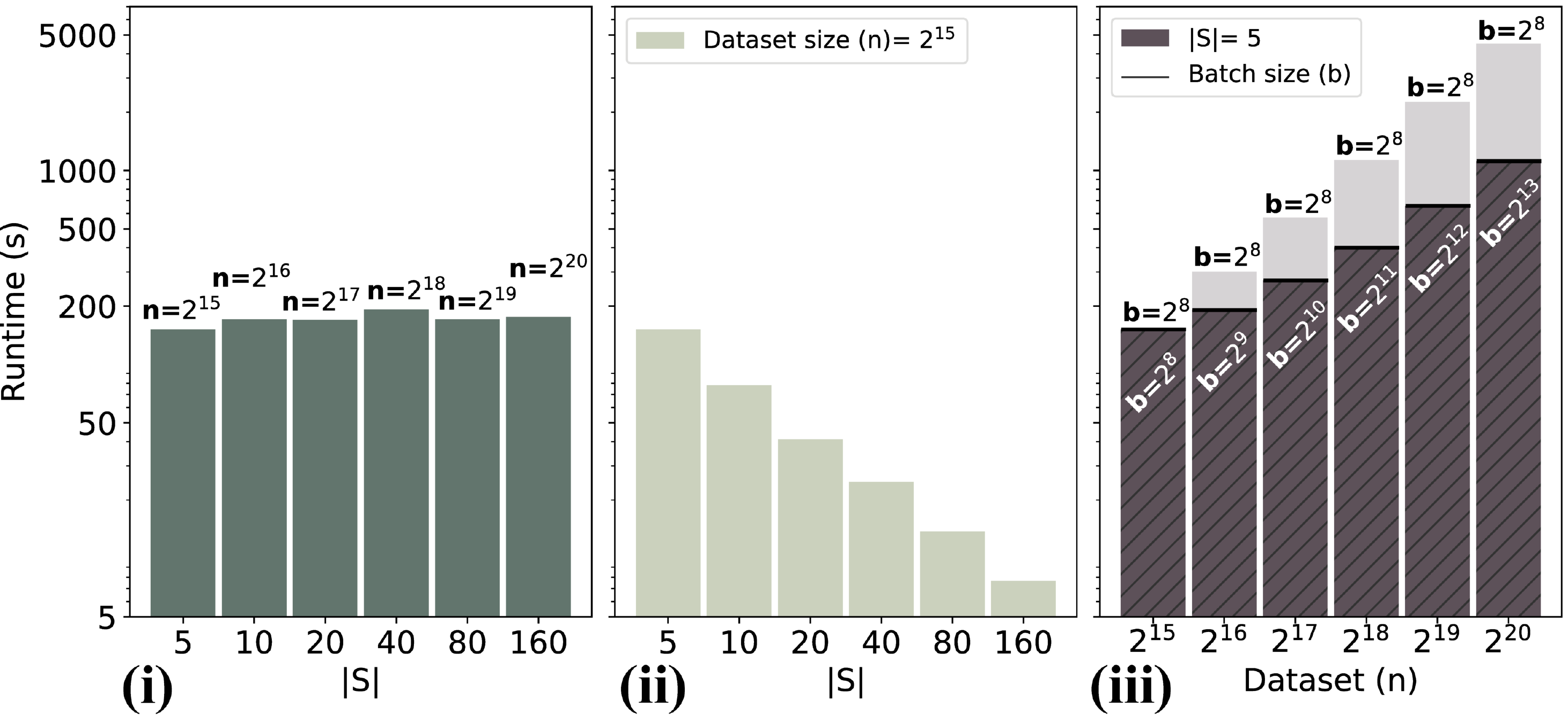}
		\caption{\sys's performance with the nbr. of DPs ($|S|$) \& records ($n$).}
		\label{fig:performance2}
	\end{subfigure}
	\caption{\centering \sys's Scalability.}
	\label{figPerformance}
\end{figure*}

\df{We employ two sets of security parameters (SP), both ensuring 128-bit security: \textsc{sp1}: ($N$ $=$ $2^{14}$, $Q$ $=$ $2^{438}$, $\eta=3.2$, number of levels $L$ $=$ $9$, scale $\Delta$ $=$ $2^{34}$, degree of the approximated activation function $d$ $=$ $5$) and \textsc{sp2}: ($N$ $=$ $2^{13}$, $Q$ $=$ $2^{218}$, $\eta=3.2$, $L$ $=$ $6$, $\Delta$ $=$ $2^{30}$, $d$ $=$ $3$). \textsc{sp2} is sufficient for linear regression and for specific logistic regression models that accept a low-degree $d$ approximation. To account for a wider-range of solutions, we rely on \textsc{sp1}, unless otherwise stated. \dfm{We employ the \textit{local bootstrap} approach for all our experiments and for all datasets in \textit{baseline comparison} except for ESR and CalCOFI, for which we use \textit{global bootstrap}, as in most cases, doing multiple $m$ local iterations between two global iterations yields a better accuracy. We further study the choice of bootstrapping strategy later in this section.} Unless otherwise stated, we employ the \textit{diagonal approach} (DA) for packing; we compare it with the \textit{row-based approach} (RBA) in Figure~\ref{fig:performanceFeature}. In all our experiments, we consider \sys's total time to train a regression model (including communication) without \textsc{prepare}, which is executed once and mostly involves light plaintext operations. E.g., the complete \textsc{prepare} takes 16.5s for a dataset of 40,000 samples distributed among 4 DPs. \textsc{map} accounts for up to $99.5\%$ of \sys's execution time. As shown in Section \ref{sec:optimizations}, the DPs perform most of the computations in \textsc{map}, which is the only step with multiple local iterations, involving two matrix-vector multiplications, which span most (up to $97\%$) of its execution time. The remaining time corresponds to the computation of activation function and collective bootstrapping.}

\begin{table}[h]
	\centering
	\tiny
	\begin{subfigure}{0.2\columnwidth}
		\centering
		\includegraphics[width=0.75\columnwidth]
		{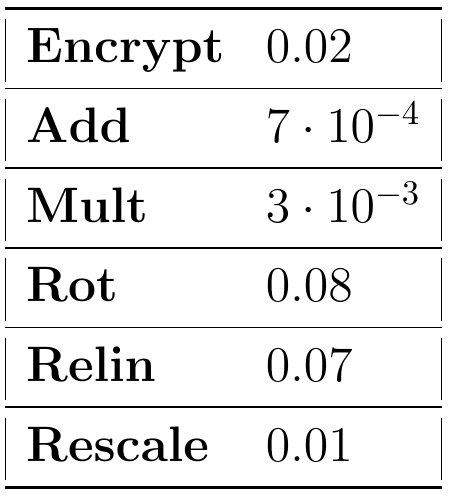}
		\vspace{-0.5em}
		\caption{\dfm{Local Crypto. Ops.}}
		\label{fig:localBenchmark}
	\end{subfigure}
	\hspace{1em}
	\begin{subfigure}{0.45\columnwidth}
		\centering
		\includegraphics[width=1.0\columnwidth]
		{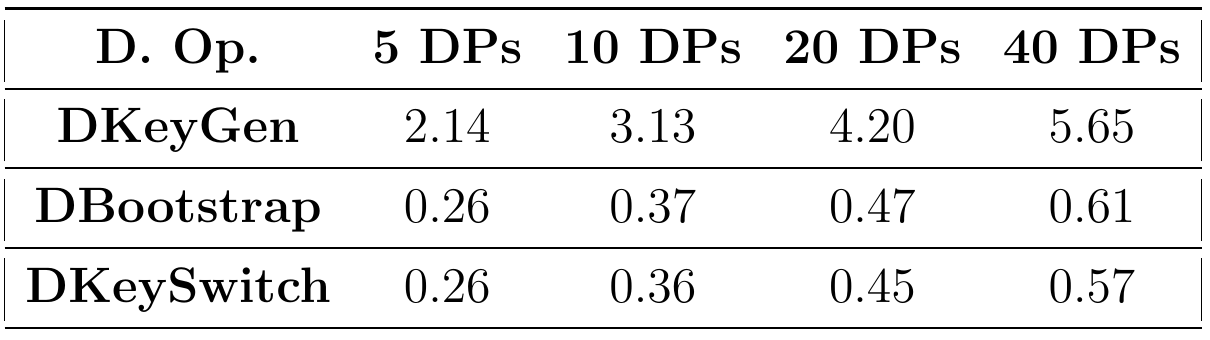}
		\vspace{-1.2em}
		\caption{\centering\dfm{Distributed Crypto Ops.}}
		\label{fig:globalBenchmark}
	\end{subfigure}
	\caption{\dfm{\centering Crypto. micro-benchmarks in seconds with \textsc{sp1}.}}
	\label{figBenchmark}
\end{table}
\dfm{\descr{Micro-benchmarks.} Table \ref{figBenchmark} shows the execution time of each cryptographic operation. We observe that, as mentioned before, \sys replaces the usually costly bootstrapping operation by an efficient interactive protocol $\text{DBootstrap}(\cdot)$. One of the most recent works on bootstrapping by Han and Ki \cite{Han_BetterBootstrap} introduces a solution that only achieves around 108 bits of security (lower than the recommended 128 bits, due to recent attacks \cite{revisitinghybridattackssparsekeys,cheon2019hybrid}) and executes a \textsc{ckks} bootstrapping in 26 seconds with ciphertexts that can encrypt $2^{13}$ values, corresponding to \textsc{sp1}, and about 20 seconds for \textsc{sp2}. This is two orders of magnitude slower than our $\text{DBootstrap}(\cdot)$, that achieves 128-bit security with execution times of 0.6 and 0.25 seconds for \textsc{sp1} and \textsc{sp2} respectively (with 40 DPs).}

\descr{Baseline Comparison.} \df{To evaluate \sys, we compare its performance (execution time and accuracy) against an \textit{ideal} baseline, i.e., a non-privacy-preserving centralized cleartext solution (CCS) where a DP obtains the full dataset and trains the model on it. We consider the training time on the complete dataset and use the training batch size $b$ as the number of data samples input for the prediction. Moreover, we compare \sys with a distributed non-privacy-preserving (DNP) solution (cleartext values and exact activation functions), to show that our cryptographic approach and activation approximations introduce minimal accuracy degradation. Finally, to demonstrate the benefit of distributed learning approaches, we compare \sys's accuracy with a case where one DP independently trains a model only on its local part of the distributed dataset (Independent Training, IT). 
\begin{table}[t]
	\vspace{-0.8em}
	\centering
	\tiny
	\includegraphics[width=0.5\columnwidth]{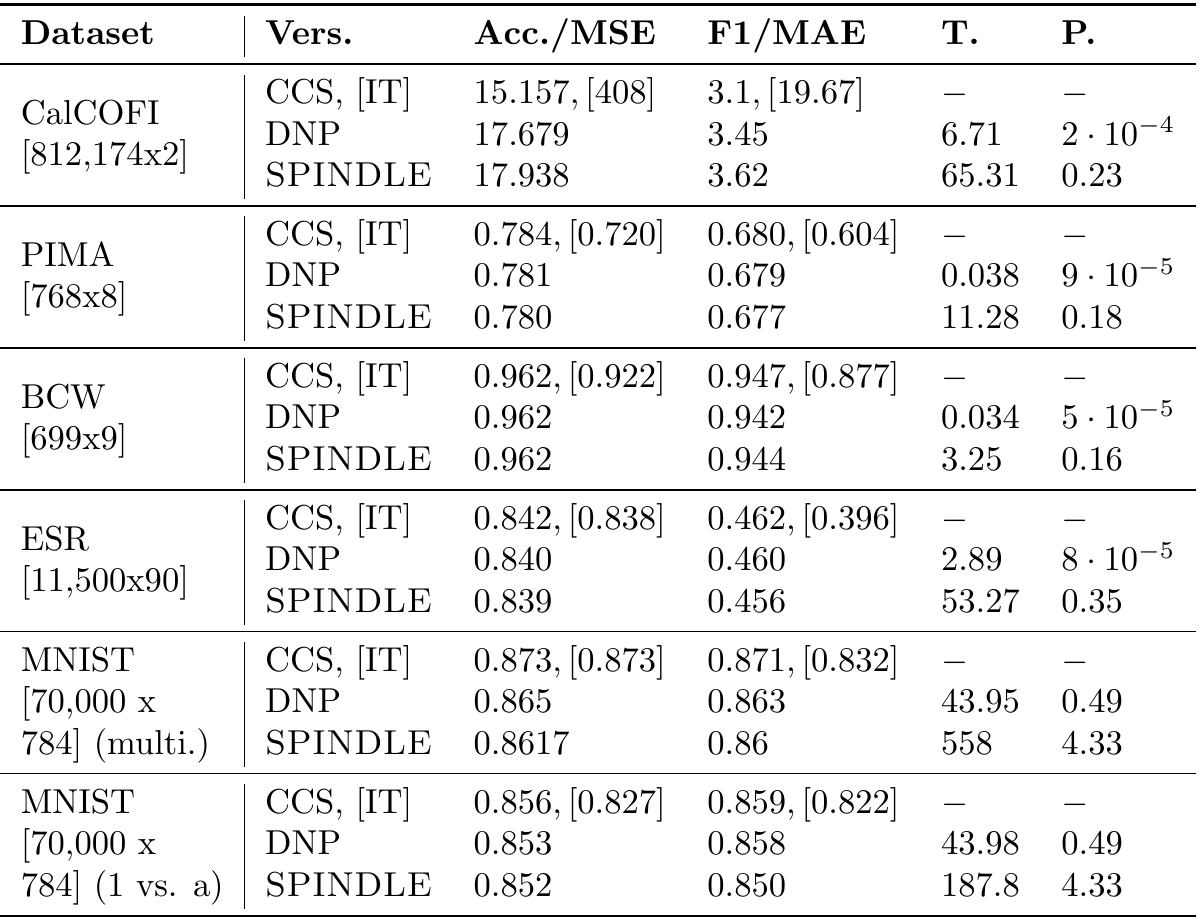}
	\caption{\small{\dfm{Baseline Comparison with K-fold=5. Time to train (T.) and to predict (P.) are in seconds. MSE and MAE are given for the lin. reg. on CalCOFI. Accuracy and F1-score are given for all the others.}}}
	\label{fig:baselineComp}
	\vspace{-3.5em}
\end{table}
In Table \ref{fig:baselineComp}, we show \sys's accuracy (Acc.) (resp., Mean Squared Error, MSE) and F1-score (F1) (resp., Mean Average Error, MAE) for logistic and multinomial (resp., linear) regressions, achieved on the above datasets when they are split among 10 DPs. We refer to Table \ref{fig:AppbaselineComp} in Appendix \ref{app:evalcomplement} for the learning parameters description. We observe that \sys's accuracy loss is very low, up to 0.8\%, with respect to a non-private centralized (CCS) solution where the model is trained on the full dataset (using standard SGD) with a standard Python library~\cite{scikit-learn}. For instance, on ESR, CCS yields 84.2\%, to 83.9\% with \sys. Moreover, this loss is mainly due to the data not being centralized, as \sys consistently achieves almost the same accuracy as the decentralized non-private (DNP) equivalent. 
\sys's total training time (column T. in Table~\ref{fig:baselineComp}) is kept between 1 and 2 orders of magnitude higher than DNP, as the costly operations on encrypted data are partially amortized by SIMD operations enabled by the used packing. For instance, the training on the ESR dataset takes almost 3 seconds in DNP and 53.27 seconds in \sys. We do not report the time for the centralized training (CCS and IT), as the settings are too different to be fairly comparable. Multinomial regression requires polynomial approximations of higher degree, i.e., between 15 and 19 (see Appendix \ref{fig:AppbaselineComp}); its training on 70,000 records of 784 features (MNIST) is executed in 558 seconds (column T. in Table~\ref{fig:baselineComp}). This time can be reduced to 187.8 seconds by performing 10 logistic regressions in parallel, one per label class (one-vs-all), at the cost of a 1\% loss in accuracy. In all cases, when a DP independently trains on its part of the dataset (IT), i.e., with $1/10$-th of the data, the achieved accuracy is worse than the one achieved on the entire distributed dataset. As for prediction (P.), \sys's prediction on 10 data samples of 90 features (ESR dataset) requires only 0.35 seconds by packing the input data and executing parallel computations. This time can be further amortized if more predictions are run in parallel.}

\descr{Scalability.} We study how \sys's execution time evolves when increasing the number of: features ($c$), data providers ($|S|$), and dataset samples ($n$). By default, we set $|S|$ $=$ $5$, each DP having $5,120$ data records (synthetically generated) with $c$ $=$ $32$ features; we use a batch size $b$ $=$ $256$, with $g=5$ global iterations, and $m=20$ local iterations in \textsc{map}. When comparing different approaches, we ensure that the number of times that the dataset is fully processed is constant, and we set the learning parameters 
accordingly. Figure~\ref{fig:performanceFeature} displays \sys's execution time with an increasing number of features $c$ and shows \df{that it scales logarithmically}, in any of the used approaches. In this setting, we also study the influence of the multi-threading, the differences between the two packing approaches (Section~\ref{sec:MatrixMultSGD}) and the impact of having encrypted input data (Section~\ref{sec:systemConfiguration}). When the computations are single-threaded, the row-based approach (RBA) is more efficient than the diagonal approach (DA) up to $c=128$ features, as RBA incurs fewer multiplications and rotations than DA. 
In contrast, the diagonal approach (DA) execution time in one or multiple threads is almost constant up to $c=256$ features (with a batch size $b=256)$, as its complexity depends mainly on $\text{max}(c,b)$ (Section~\ref{sec:MatrixMultSGD}).  However, the DA is \textit{embarrassingly parallelizable}, and it is always faster when the computations are executed on 24 threads. As an example, on multiple threads and for 256 features, DA yields an execution time of 165s against 330s for RBA, and 365s when the input data are encrypted and using RBA. For both approaches, the parallelization is efficient up to $c=2^8$, where the maximum thread-utilisation is reached. Afterwards, both approaches scale linearly. When the data providers have encrypted input data (RBA-E), the execution time increases by $7\%$ with respect to RBA.

Figure~\ref{fig:performance2}.i shows that when the number of DPs $|S|$ increases and each DP has a fixed amount of data, \sys's execution time is constant. This means that \sys scales independently of $|S|$. In Figure~\ref{fig:performance2}.ii, where $|S|$ increases but the total amount of data remains constant, \sys's execution time decreases linearly, as the workload is efficiently distributed among the DPs. In Figure \ref{fig:performance2}.iii, when $|S|$ is constant and the size of the DPs' datasets increases, \sys's execution time increases linearly with the amount of data. If the batch size can be increased when the data providers have more records, then \sys's execution time can be further reduced. In summary, \sys scales independently of the number of data providers, and linearly with the DPs' dataset size. It is able to train models with a high number of features and thus remains practical for real-world sized datasets. \dfm{Finally, we note that \sys scales similarly to a DNP solution in the three cases. For example, in the case of Figure~\ref{fig:performance2}.ii, DNP ranges from 0.69 seconds with 5 DPs to 0.42 seconds with 10 DPs, whereas \sys's execution time decreases from 150 to 78 seconds.}

\dfm{\descr{Bootstrapping \& Activation Function Approx. Degree.}}
\dfm{In Table \ref{fig:bootActiv}, we observe that \sys accuracy slightly improves with higher degree approximations of the sigmoid for the activation function. Relying on LB or HB, which require less global iterations and therefore less communication, and on low-degree approximations improves \sys execution time but, it can lower the achieved accuracy. We remark that HB requires less bootstrapping operations, as the global weights are bootstrapped and assigned to the local weights. However, LB and HB execution times remain similar as in LB, the DP perform the bootstrappings in parallel.}
\begin{table}[h]
	\centering
	\tiny
	\includegraphics[width=0.35\columnwidth]{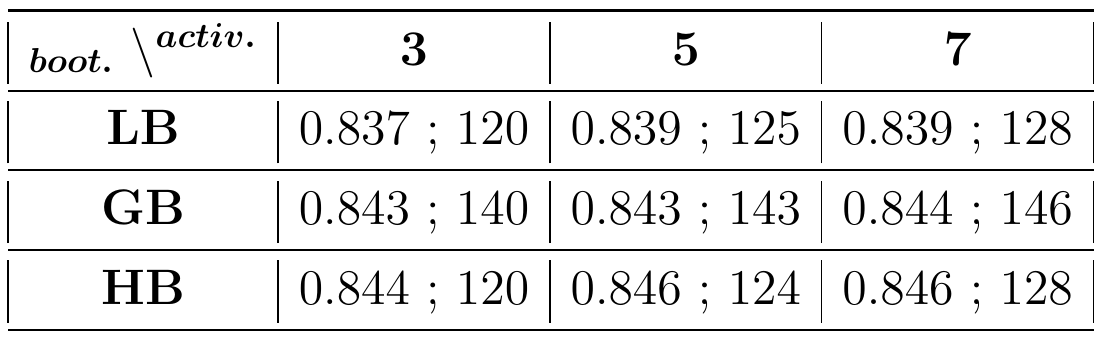}
	\caption{\small{\dfm{\sys accuracy and timing \textit{(accuracy;exec. time in sec)} to train on ESR with different bootstrapping strategies and degrees of the activation function.}}}
	\label{fig:bootActiv}
\end{table}

\descr{Communication.} \df{With the security parameters \textsc{sp1}, the size of a ciphertext is 2.6MB and each DP receives and sends one ciphertext per global iteration. One ciphertext is also exchanged for each $\text{DBootstrap}(\cdot)$ (e.g., every two global iterations).}

\descr{Comparison with prior art.}
\df{Here we briefly compare, both qualitatively and quantitatively (when applicable), \sys against (a) centralized cryptographic approaches (e.g., ~\cite{kim2018secure,kim2018logistic,carpov2019privacy}), (b) cryptographic distributed solutions (e.g., ~\cite{Drynx, corrigan2017prio,zheng2019helen}) and (c) federated learning solutions (e.g., \cite{du2018privacy, huang2019dp, Nvidia_Fed, shokri2015privacy}). See Appendix \ref{app:compareprior} for an extended analysis.}

\descr{(a)} \df{\sys consistently outperforms centralized HE-based solutions (CES) as \sys distributes the workload among multiple DPs and replaces, \dfm{as shown before}, the costly centralized bootstrapping operation by a lightweight interactive protocol.} 

\descr{(b)} \df{None of the existing HE-SMC-based distributed solutions~\cite{Drynx, corrigan2017prio,zheng2019helen} provides both data and model confidentiality, or covers the entire ML workflow (the trained model cannot be kept secret to perform oblivious predictions) or enables the distributed execution of the gradient descent. Moreover, some solutions (\cite{Drynx, corrigan2017prio}) leak more than only the trained model and rely on data encodings (or approximations) that lower the obtained accuracy, whereas in \sys, we approximate only the activation functions. Finally, all previous solutions scale quadratically in at least one dimension, i.e., number of features $c$, samples $n$, or DPs $|S|$, whereas \sys's execution time is almost independent of $|S|$, scales logarithmically with the number of features and linearly with the dataset size. Purely secret-sharing-based solutions \cite{bogdanov2016rmind, Cho_GWAS} consider substantially different settings as \sys, as they require the DPs to communicate their data outside their premises and require an honest majority among a limited number of computing servers (typically, 2 to 4, depending on the setting).} \dfm{Whereas \sys also works in this configuration, it is not specifically optimized for 2-4 parties and its execution time would be in the same order of magnitude but slower than secret-sharing-based solutions. This is due to the computation overhead introduced by operations on encrypted data. However, unlike secret-sharing-based solutions, \sys efficiently scales to federated learning settings where many (hundreds of) DPs keep their data locally and can withstand up to N-1 out of N dishonest DPs.}

\dfm{\descr{(c)} In basic federated learning solutions, data owners train and update the model on their local data and a server aggregates the model updates to obtain the global model~\cite{federatedLearning1,Konency2016fed}. In this setting, the coordinating server has to be fully trusted, as some information can be inferred from the intermediate models, e.g., extracting participants' inputs~\cite{hitaj2017deep,Wang2019,NIPS2019_9617} or membership inference~\cite{Melis2019,Nasr2019}. \sys naturally thwarts federated-learning and model-inversion attacks, as the intermediate and final weights are never revealed.} Federated learning approaches based on differential privacy (diffP), e.g., ~\cite{Nvidia_Fed,shokri2015privacy,McMahan2018}, train the model while introducing noise to the intermediate values to mitigate adversarial inferences. These approaches consider a different paradigm by introducing a tradeoff between privacy and accuracy, whereas in \sys security is absolute, and the trade-off (accuracy vs execution time) is the same as for non-secure solutions, e.g., less training iterations can yield a less precise model. DiffP approaches can significantly degrade the data utility, and might require a high privacy budget for which it remains unclear what privacy protection is obtained in practice~\cite{jayaraman2019evaluating}. \dfm{In Section \ref{sec:extensions}, we discuss how membership inference and reconstruction attacks from the prediction outputs can also be mitigated in \sys by adding differentially-private noise during the $\text{DKeySwitch}(\cdot)$.}

%% file: discussion.tex
\section{Extensions}\label{sec:extensions}

\dfm{We describe here extensions to \sys that can be employed to withstand  malicious adversaries, support dynamic DPs, enable quality control and support more complex ML models.} 

\dfm{\subsection{Malicious Adversaries}} \label{extension_malicious}

\dfm{\descr{Malicious DPs Interfering with \textsc{Training}.} To limit the extent to which a malicious DP could interfere with its \textsc{training}, \sys can require from the DPs to publish transcripts of their computations~\cite{Drynx} and to produce proofs of correct inputs. These features combined would enable \sys to be fully auditable. Mechanisms to avoid model poisoning attacks when the input data are encrypted (and have to remain confidential) are an open research problem. However, \sys can partially mitigate this threat by constraining the DPs' inputs and requiring zero knowledge proofs of range~\cite{libert2018lattice,yang2019efficient, baum2016efficient, baum2020concretely} from the DPs. This would substantially limit the extent to which a malicious DP could interfere with \sys's \textsc{training}. However, we note that this does not thwart all possible attacks, as, for example, poisoning attacks would still be possible with plausible, i.e., in-range, input data.}

\dfm{\descr{Malicious DPs Interfering with \textsc{Prediction}.} As for the \textsc{training}, computation correctness can be verified through computations' transcripts published by the DPs. To prevent a malicious DP from learning a victim querier's prediction outputs via a replay attack (i.e., reusing the querier's encrypted data in a new query), \sys can require queriers to provide signed proofs of knowledge of the input data~\cite{lyubashevsky2020practical}.}

\dfm{\descr{Malicious Querier Inferring Information from \textsc{Prediction}'s Output.} \sys naturally covers federated learning attacks~\cite{hitaj2017deep, Melis2019, Nasr2019} and model inversion attacks~\cite{fredrikson2015model}, as the intermediate and final weights are never revealed. Moreover, \sys can also mitigate inference attacks, e.g., membership inference~\cite{shokri2017membership}, by limiting the number of prediction requests on the trained model. This solution can be improved by adding noise to the prediction output to achieve differential privacy guarantees. In fact, a mechanism that ensures differential privacy can be used for all the outputs of \sys: on the predictions $\bm{y'}$ and on the trained model, if it is released after training (Section~\ref{sec:systemConfiguration}). This would ensure that a passive adversary (e.g., trying to infer information from the system's outputs) or an active adversary controlling a subset of the DPs cannot learn information, e.g., data or local model of honest parties, about a subset of the DPs. To ensure differential privacy, \sys should add some collectively generated noise~\cite{Unlynx,kim2020} to the query result before performing $\text{DKeySwitch}(\cdot)$. However, the choice of the privacy parameters is not trivial and is an interesting direction for future work. Furthermore, the use of differential privacy in dynamic systems presents serious limitations; minimizing the released non-encrypted information (which also reduces the noise magnitude required to meet a target differential privacy level) is much more effective and practical. This is the approach taken in \sys, contrarily to federated learning systems, where the intermediate outputs of each training iteration are always disclosed.}

\dfm{\subsection{Modular Extensions}}

We discuss here a set of extensions that can be (optionally) integrated and combined in \sys depending on the application.

\dfm{\descr{Threshold-encryption Scheme.} 
To account for unresponsive DPs, \sys can use a threshold-encryption scheme, where the DPs secret-share~\cite{shamir1979share} their secret keys, thus enabling a subset of the DPs to perform the cryptographic interactive protocols ($\text{DBootstrap}(\cdot)$ or $\text{DKeySwitch}(\cdot)$).}

\dfm{\descr{Dynamic Roles.} 
The role of $DP_R$ played by one DP has no security implications and only incurs small computation overhead for one DP. This role can be dynamically assigned (e.g., round robin) at each global iteration or whenever the DP playing $DP_R$ becomes unavailable.}

\dfm{\descr{Asynchronous Learning \& Performance Optimizations.} \df{We experimentally observed that an uneven distribution of the data across DPs does not affect the training accuracy. However, and as expected, in order to obtain similar accuracy as a centrally trained model, the labels of the DPs' local datasets should be similarly distributed. For this, \sys can integrate optimizations of the stochastic gradient descent (SGD) that can be expressed as a polynomial; in particular, SGD asynchronous variants that account for imbalances in DPs' response times or data distribution, or for sparse networks adaptations (e.g., Koloskova et al.~\cite{Jaggi_2019}).}

To avoid over- or under-fitting, which often happens when the number of training iterations is predefined, \sys can integrate a collective stop-test protocol. This protocol enables the data providers to collectively decrypt the absolute difference between the (global) weights of two subsequent (global) iterations, or a statistic derived from these values. The decrypted value is compared to a chosen threshold to stop the training.}

\dfm{\descr{Data Preparation \& Quality Control.}}
\df{ As mentioned before, the training on a distributed dataset can be optimized according to how the data are distributed among the DPs. This information can also serve for data standardization and quality control, and its leakage can be mitigated by relying on differential privacy or on HE-based interactive protocols \cite{Drynx}. \sys's \textsc{prepare} phase can be extended to include these solutions and it is up to the DPs to choose the configuration that achieves the required balance between privacy and performance.}

\dfm{\subsection{More Complex ML Models.}}

\dfm{We first remark that the extended, privacy-preserving MapReduce abstraction on which we rely to build \sys can actually capture many of existing solutions for secure distributed ML training \cite{shokri2015privacy,Drynx,corrigan2017prio,zheng2019helen,Bonawitz2019,nikolaenko2013privacy,giacomelli2018privacy,mohassel2017secureml,bogdanov2016rmind,Cho_GWAS}. We also remark that, even though we rely on the widely applicable distributed stochastic gradient descent (SGD), other distributed approaches for training ML models such as ADMM~\cite{boyd2011distributed} could also be expressed in the same abstraction. However, by relying on SGD, we aim at designing a system that can then be extended to other models, as SGD can be used to minimize many cost functions \cite{Kumar_2015, toulis2014statistical, Zhang_2004}. In particular, it can be extended to more complex models such as neural networks, which are usually trained using SGD~\cite{NN_SGD}. 
\sys supports any activation function that can be “practically” approximated by a polynomial; hence, the challenges for its extension to more complex models reside in trading-off precision for efficiency when approximating non-polynomial functions, and efficiently packing the data depending on the operations. This is particularly important for neural networks in which the computations are sequentially performed through multiple layers. Thus, each SGD iteration would involve higher multiplicative-depth circuits and their evaluation under encryption.} 

%% file: conclusion.tex
\section{Conclusion}\label{sec:conclusion}

By extending the MapReduce abstraction, we have proposed a generic solution to the problem of privacy-preserving distributed ML model training and prediction. Our abstraction enables us to optimize the application of protection primitives from multiparty homomorphic encryption in a MapReduce workflow. We \df{proposed \sys, a privacy-preserving system that enables the execution of a distributed stochastic gradient descent and we have instantiated our quantum-resistant solution for the training and oblivious prediction on generalized linear models}. We have shown that \sys achieves accuracy comparable to non-secure centralized solutions, and it scales independently of the number of DPs and linearly or better with the size of the DPs' local datasets \df{and the number of features. This makes it particularly suitable for difficult and demanding learning tasks that have to be performed on sensitive data that cannot be shared. This is the case in many domains and particularly in medicine, where complex sensitive datasets partitioned across medical institutions need to be regularly analyzed, e.g., Genome Wide Association Studies}. \sys achieves better performance than existing centralized and distributed solutions by leveraging the data providers concurrent computation on their local data, and using a multiparty encryption scheme that replaces costly homomorphic operations (e.g., bootstrapping) by efficient collective protocols. To the best of our knowledge, \sys is the first highly scalable system enabling the distributed execution of the gradient descent across hundreds of parties and large datasets in a privacy-preserving, post-quantum, and efficient way.

\section*{Acknowledgment}
We would like to thank all of those who reviewed the
manuscript, in particular: Henry Corrigan-Gibbs, the members of the EPFL Laboratory for Data Security and the anonymous reviewers. We also thank Yupeng Zhang for shepherding the paper. This work was partially supported by the grant \#2017-201 of the Strategic Focal Area “Personalized Health and Related Technologies (PHRT)” of the ETH Domain.

%% file: appendix.tex
\appendices
\section{Multiparty Homomorphic Encryption (MHE)}\label{app:mhe}
We describe here the cryptographic operations and distributed protocols that are used in \sys.

\descrit{Operations:} In Scheme~\ref{fig:ckksScheme}, we introduce \textsc{ckks} operations. $\bm{v}$ is a vector of cleartext values, $sk$ and $pk$ are the secret and public keys, and $evk$ is an evaluation key.

\begin{scheme}
\begin{center}
\begin{small}
\noindent\rule{\columnwidth}{0.8pt}
$\begin{array}{rl}
    \text{Encrypt:}  & \langle \bm{v}\rangle_{pk} = \{ \{\langle\bm{v}\rangle, \tau, \Delta\}, L, \Delta\}_{pk} = \text{Enc}(pk,\bm{v})\\
    \text{Decrypt:} & \bm{v} = \text{Dec}(sk,\langle\bm{v}\rangle_{pk}) \\
    \text{Add:}    & \{\langle\bm{v}_1\rangle\;+\;\langle\bm{v}_2\rangle, \min(\tau, \tau'),\max(\Delta, \Delta')\}\; =  \\
                         & \{\langle\bm{v}_1\rangle, \tau,\Delta\}\; +\; \{\langle\bm{v}_2\rangle, \tau',\Delta'\} \\
    \text{Mult:} & \{\langle\bm{v}_3\rangle,\min(\tau, \tau'),\Delta\Delta'\}=\text{M}(\{\langle\bm{v}_1\rangle,\tau,\Delta\}\;,\; \{\langle\bm{v}_2\rangle,\tau',\Delta'\}) \\
    \text{Rot:} & \{\langle\bm{v'}\rangle,\tau,\Delta\}\; =\; \text{RotL/R}(\{\langle\bm{v}\rangle,\tau,\Delta\}, r, evk)\\
    \text{Rescale:} & \{\langle\bm{v}\rangle,\tau-1,\Delta'\}\; =\; \text{ReScale}(\{\langle\bm{v}\rangle,\tau,\Delta\})\\
    \text{Relin:} & \{\langle\bm{v}\rangle,\tau,\Delta\}\; =\; \text{Relin}(\{\langle\bm{v}\rangle,\tau,\Delta\}, evk)\\
    \text{Bootstrap:} & \{\langle\bm{v}\rangle,\tau,\Delta\}\;=\; \text{Bootstrap}(\{\langle\bm{v}\rangle,0,\Delta\}, evk)\\
\end{array}$
\noindent\rule{\columnwidth}{0.8pt}
\end{small}
\end{center}
\caption{\centering \textsc{CKKS} operations.}
\label{fig:ckksScheme}
\end{scheme}

\descrit{Distributed Protocols:} In Scheme \ref{fig:distribscheme}, we define four protocols using the distributed version of \textsc{ckks}. These protocols require the participation of all DPs, i.e., each $DP_i$ contributes its respective secret key $sk_i$. $\text{DKeyGen}(\cdot)$ generates the collective public key $pk$ and evaluation keys $evks$ by combining the protocols defined by Mouchet et al.~\cite{mouchet2019distributedbfv}. These keys can then be used independently (without interaction) by a DP on a ciphertext $\langle\bm{v}\rangle_{pk}$. The distribution of the scheme enables an efficient distributed bootstrapping $\text{DBootstrap}({\langle\bm{v}\rangle,\tau_b,\Delta}, \{sk_i\})$ to collective refresh $\langle\bm{v}\rangle$ to its initial level $L$. The minimum level $\tau_b$ at which the bootstrapping has to be performed depends on the security parameters. The $\text{DKeySwitch}(\cdot)$ enables the DPs to change a ciphertext encryption from the public key $pk$ to another public key $pk'$, without decrypting the ciphertext. A distributed decryption operation $\text{DDec}(\cdot)$ is a special case of the $\text{DKeySwitch}(\cdot)$ where there is no $pk'$, i.e., $pk'$ is 0.
\begin{scheme}
\begin{center}
\begin{small}
\noindent\rule{\columnwidth}{0.8pt}
$\begin{array}{rl}
    \text{Distrib. Key Gen:}  & pk, evks\;=\;\text{DKeyGen}(\{sk_i\})\\
    \text{Distrib. Bootstrap:} & \{\langle\bm{v}\rangle,L,\Delta\} = \text{DBootstrap}({\langle\bm{v}\rangle,\tau_b,\Delta}, \{sk_i\})\\
    \text{Distrib. Key Switch:} & \langle\bm{v}\rangle_{pk'}\;=\;  \text{DKeySwitch}(\langle\bm{v}\rangle_{pk}, pk', \{sk_i\}) \\
    \text{Distrib. Decrypt:} & \bm{v} = \text{DDec}(\langle\bm{v}\rangle, \{sk_i\})\\
\end{array}$
\noindent\rule{\columnwidth}{0.8pt}
\end{small}
\end{center}
\caption{\centering Distributed \textsc{CKKS} operations.}
\label{fig:distribscheme}
\end{scheme}

\section{Activation Functions}\label{app:OptimizedActivation}
We describe how we evaluate a polynomial approximation and how we approximate the maximum function.

\descrit{Polynomial Approximation}
Protocol~\ref{alg:OptimizedActivation} inductively computes the (element-wise) exponentiation of the encrypted input vector $\langle\bm{u}\rangle$: $\langle\bm{u}^{1}\rangle$, $\langle\bm{u}^{2}\rangle$, $\dots$, $\langle\bm{u}^{2^k-1}\rangle$, $\langle\bm{u}^{2^k}\rangle$, $\langle\bm{u}^{2^{k+1}}\rangle$, $\dots$, $\langle\bm{u}^{2^{\omega-1}}\rangle$ (Protocol~\ref{alg:OptimizedActivation}, line 2), where $\omega$ is the smallest value satisfying $2^{\omega}>d(p(\langle\bm{u}\rangle))$ and $k=\lfloor \omega/2\rfloor$. Then, it recursively evaluates $p(\langle\bm{u}\rangle) = \sum_{i=1,2,3...,d}r_i \langle\bm{u}^{i}\rangle = \langle\bm{u}^{2^{\omega-1}}\rangle q(\langle\bm{u}\rangle)+R(\langle\bm{u}\rangle)$ (Protocol~\ref{alg:OptimizedActivation}, line 3). Note that $p(\cdot)$, $q(\cdot)$, and $R(\cdot)$ are functions of $\langle\bm{u}\rangle$ and of the approximation coefficients $\bm{r}$, $q(\cdot)$ is the quotient of the division of the actual activation function $p(\cdot)$ by $\langle\bm{u}^{2^{\omega-1}}\rangle$, and $R(\cdot)$ is the remainder of the division. $d(x)$ is a function that outputs the degree of $x$. 
\begin{figure}[h]
\begin{protocol}[H]
    \small
	\caption{Encrypted Poly. Approx. Evaluation $\text{AF}(\cdot)$.}
	\label{alg:OptimizedActivation}
	\renewcommand{\thealgorithm}{}
	\begin{algorithmic}[1]
	\item[Func. $\text{AF}(\langle \bm{u} \rangle,d,\bm{r})$ outputs $\langle \bm{a} \rangle$ the evaluated poly. approx. of $\langle \bm{u} \rangle$]
	    \STATE Choose the smallest $\omega$ such that $2^\omega>d$ and define $k=\lfloor \omega/2 \rfloor$
	    \STATE Compute $\{u_i\}=$ $\langle\bm{u}^{1}\rangle,\langle\bm{u}^{2}\rangle, \dots, \langle\bm{u}^{2^k\text{-}1}\rangle, \langle\bm{u}^{2^k}\rangle, \langle\bm{u}^{2^{k\text{+}1}}\rangle,\dots,\langle\bm{u}^{2^{\omega\text{-}1}}\rangle$
	    \sblinesmall
	    inductively and call $\text{\textbf{paRecu}}(\bm{r}, d, \{u_i\}$)
	    \sbline
	    \STATE \textbf{Function} \text{\textbf{paRecu}}$(\bm{r}, d, \{u_i\}$)\textbf{:}
	    \STATE \quad Choose the smallest $\omega$ such that $2^\omega>d$
	    \sblinesmall
	    \STATE \quad Find polynomials $q(\langle\bm{u}\rangle)$ and $R(\langle\bm{u}\rangle)$ with $\langle\bm{a}\rangle=$\\
	    \quad $ \langle\bm{u}^{2^{\omega-1}}\rangle q(\langle\bm{u}\rangle)+R(\langle\bm{u}\rangle)$ such that $\langle\bm{a}\rangle = {\sum_{i=1,2,..,d}}\bm{r}[i] \langle\bm{u}^{i}\rangle$
	    \sblinesmall
	    \STATE \quad \textbf{If} $d(q),d(R)\leq 2$ \textbf{:}\\ 
	    \quad \quad Evaluate $q(\langle\bm{u}\rangle)=\text{\textbf{paRecu}}(\bm{r}, d = d(q), \{u_i\})$\\ 
	    \quad \quad and $R(\langle\bm{u}\rangle)=\text{\textbf{paRecu}}(\bm{r}, d = d(R), \{u_i\})$ \sblinesmall
	    \STATE \quad \textbf{Else} Return $\langle\bm{a}\rangle$
	    \vspace{-0.3em}
	\end{algorithmic}
\end{protocol}
\captionsetup{labelformat=empty}
\end{figure}

\descrit{Approximation of the maximum function}
Protocol (Protocol~\ref{alg:max}) computes the approximation of the maximum function. 
It takes an encrypted matrix $\langle\bm{U}_{|cl| \times c}\rangle$, the approximations intervals  $[a_i,g_i]$ and degrees $\bm{d}$, and computes an encrypted vector $\langle\bm{m}\rangle$ that contains a close approximation of the max of each column of $\langle\bm{U}\rangle$.
\begin{figure}[h]
\begin{protocol}[H]
    \small
	\caption{Approximation of the max function $\text{apMax}(\cdot)$.}
	\label{alg:max}
	\renewcommand{\thealgorithm}{}
	\begin{algorithmic}[1]
	    \item[]{}\hspace{-0.4cm}$\langle\bm{m}\rangle \leftarrow \text{apMax}(\langle\bm{U}\rangle, [a_i,g_i], \bm{d})$
	    \STATE $\langle\bm{u'}\rangle = \sum_{\lambda=0}^{|cl|} \langle\bm{U}[\lambda,\cdot]\rangle$
	    \sblinesmall
	    \STATE \textbf{for} $\lambda=1,\dots,|cl|$\textbf{:} $\langle\bm{U}[\lambda,\cdot]\rangle = (\langle\bm{U}[\lambda,\cdot]\rangle -\langle\bm{u'}[\lambda,\cdot]\rangle)$  
	    \sblinesmall
	    \STATE $\bm{r} \leftarrow \text{GetAFCoefficients}((1/h')e^{(x/h)},[a_1,g_1],\bm{d}[1])$, where $h,h'$ are predefined constants 
	    \STATE \textbf{for} $\lambda=1,\dots,|cl|$\textbf{:} $\langle\bm{U''}[\lambda,\cdot]\rangle = \text{AF}(\langle\bm{U}[\lambda,\cdot]\rangle, \bm{d}[1], \bm{r})$
	    \STATE $\langle\bm{o}\rangle = \sum_{\lambda=0}^{|cl|} \langle\bm{U''}[\lambda,\cdot]\rangle$
	    \sblinesmall
	    \STATE $\bm{r'} \leftarrow  \text{GetAFCoefficients}(\{1/x\}, [a_2,g_2], \bm{d}[2])$ 
	    \STATE $\langle\bm{o}\rangle = \text{AF}(\langle\bm{o}\rangle, \bm{d}[2], \bm{r'})$
	    \sblinesmall
	    \STATE \textbf{for} $\lambda=1,\dots,|cl|$\textbf{:} $\langle\bm{U}[\lambda,\cdot]\rangle = \text{M}(\langle\bm{U}[\lambda,\cdot]\rangle,\langle\bm{U''}[\lambda,\cdot]\rangle)$
	    \sblinesmall
	    \STATE $\langle\bm{m}\rangle = \sum_{\lambda=0}^{|cl|}(\langle\bm{U}[\lambda,\cdot]\rangle,\langle\bm{o}\rangle)$
	\end{algorithmic}
\end{protocol}
\end{figure}

\section{Security of DBootstrap$(\cdot)$}\label{app:DBootstrapSecurity}

The original distributed bootstrapping protocol for \textsc{BFV} \cite{fan2012somewhat} is presented by Mouchet et al. \cite{mouchet2019distributedbfv}. In this protocol, the data providers produce an
additive sharing of the encrypted ciphertext by masking their share in the decryption, before collectively encrypting their share to collectively produce a new (fresh) encryption of the same value. We adapted this protocol to the \textsc{ckks} scheme \cite{cheon2017homomorphic}. The protocol steps remain the same but the underlying computational assumptions are different. In fact, in \textsc{ckks} the shares created by the data providers are not unconditionally hiding, but statistically or computationally hiding due to the incomplete support of the used masks. The proof for the protocol's \textsc{ckks} version ($\textsf{DBootstrap}(\cdot)$) follows from the proof provided by Mouchet et al. in the passive-adversary model of the BFV bootstrapping protocol with the additional assumption that Lemma~\ref{lemmaDBoot} is true. This lemma guarantees the statistical indistinguishablity of the shares in $\mathbb{C}$. The \textsc{RLWE} problem is hard if the adversary is computationally-bounded, whereas Lemma~\ref{lemmaDBoot} relies on a statistical argument. However, both share the same security bound given the same security parameter and $\textsf{DBootstrap}(\cdot)$ provides the same computational security as Mouchet et al. \cite{mouchet2019distributedbfv} original protocol.

\begin{lemma}
Given the distribution $P_{0} = (a+b)$ and $P_{1} = c$ with $0 \leq a < 2^{\delta}$ and $0 \leq b, c < 2^{\lambda + \delta}$ and $b$, $c$ uniform, then the distributions $P_{0}$ and $P_{1}$ are $\lambda$-indistinguishable; i.e., a probabilistic polynomial adversary $\mathcal{A}$ cannot distinguish between both with probability greater than $2^{-\lambda}$: $|\text{Pr}[\mathcal{A}\rightarrow 1 | P = P_{1}] - \text{Pr}[\mathcal{A}\rightarrow1|P = P_{0}]|\leq 2^{-\lambda}$.
\label{lemmaDBoot}
\end{lemma}

\descr{Proof:} We refer to Algesheimer et. al~\cite{Algesheimer2002}, Section 3.2 and Schoenmakers and Tuyls~\cite{Schoenmakers2006}, Appendix A, for the proof of the statistical $\lambda$-indistinguishability.


We recall that an encoded message $msg$ of $N/2$ complex numbers with the CKKS scheme is an integer polynomial of $\mathbb{Z}[X]/(X^{N}+1)$. Given that $||msg|| < 2^{\delta}$, and a second polynomial $M$ of $N$ integer coefficients with each coefficient uniformly sampled and bounded by $2^{\lambda +\delta} -1$ for a security parameter $\lambda$, Lemma~\ref{lemmaDBoot} suggests that Pr$[||msg^{(i)} + M^{(i)}|| \geq 2^{\lambda + \delta}] \leq 2^{-\lambda}$, for $0 \leq i < N$ and where $i$ denotes the $i^{th}$ coefficient of the polynomial.
That is, the probability of a coefficient of $msg + M$ to be distinguished from a uniformly sampled integer in $[0, 2^{\lambda + \delta})$ is bounded by $2^{-\lambda}$. 
In Mouchet et al. protocol, each party samples its polynomial mask $M$ with uniform coefficients in $[0, 2^{\lambda + \delta})$.

The parties, however, should have an estimate of the magnitude of $msg$ to derive $\delta$, which can be derived from the plaintext scale, integer precision and previous homomorphic operations. The masks $M_{i}$ are added to the ciphertext of $R_{Q_{\ell}}$ during the switch to the secret-shared domain.
To avoid a modular reduction of the masks in $R_{Q_{\ell}}$ and ensure a correct re-encryption in $R_{Q_{L}}$, the modulus $Q_{\ell}$ should be large enough for the additions of $N$ masks. 
\section{Evaluation} \label{app:evalcomplement}
We first provide the complete complexity analysis of \sys. We describe in more details the datasets used to assess \sys's training accuracy and show in Table \ref{fig:AppbaselineComp} an extended version of Table \ref{fig:baselineComp} including the learning and approximation parameters.

\label{app:complexity}
\begin{table*}[h]
	\centering
	\tiny
	\begin{subfigure}[t]{0.42\textwidth}
		\centering
	    \raisebox{2.5pt}{\includegraphics[width=1.0\columnwidth]{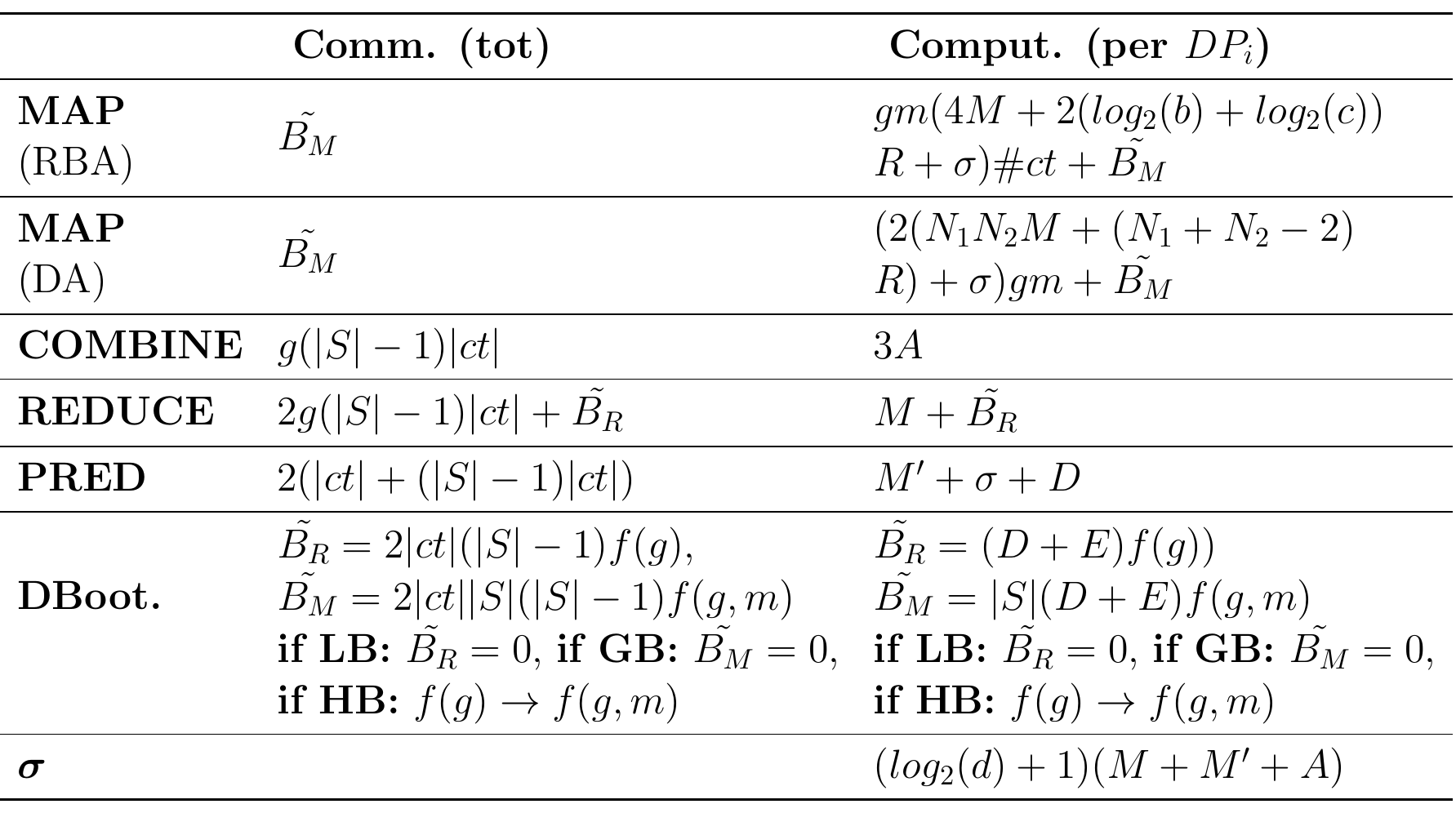}}
		\caption{\dfm{Theoretical Analysis. The number of HB always depends on both $g$ and $m$.}}
		\label{fig:theoryAnalysis}
	\end{subfigure}
\hspace{1em}
	\begin{subfigure}[t]{0.55\textwidth}
		\centering
		\includegraphics[width=1\columnwidth]
		{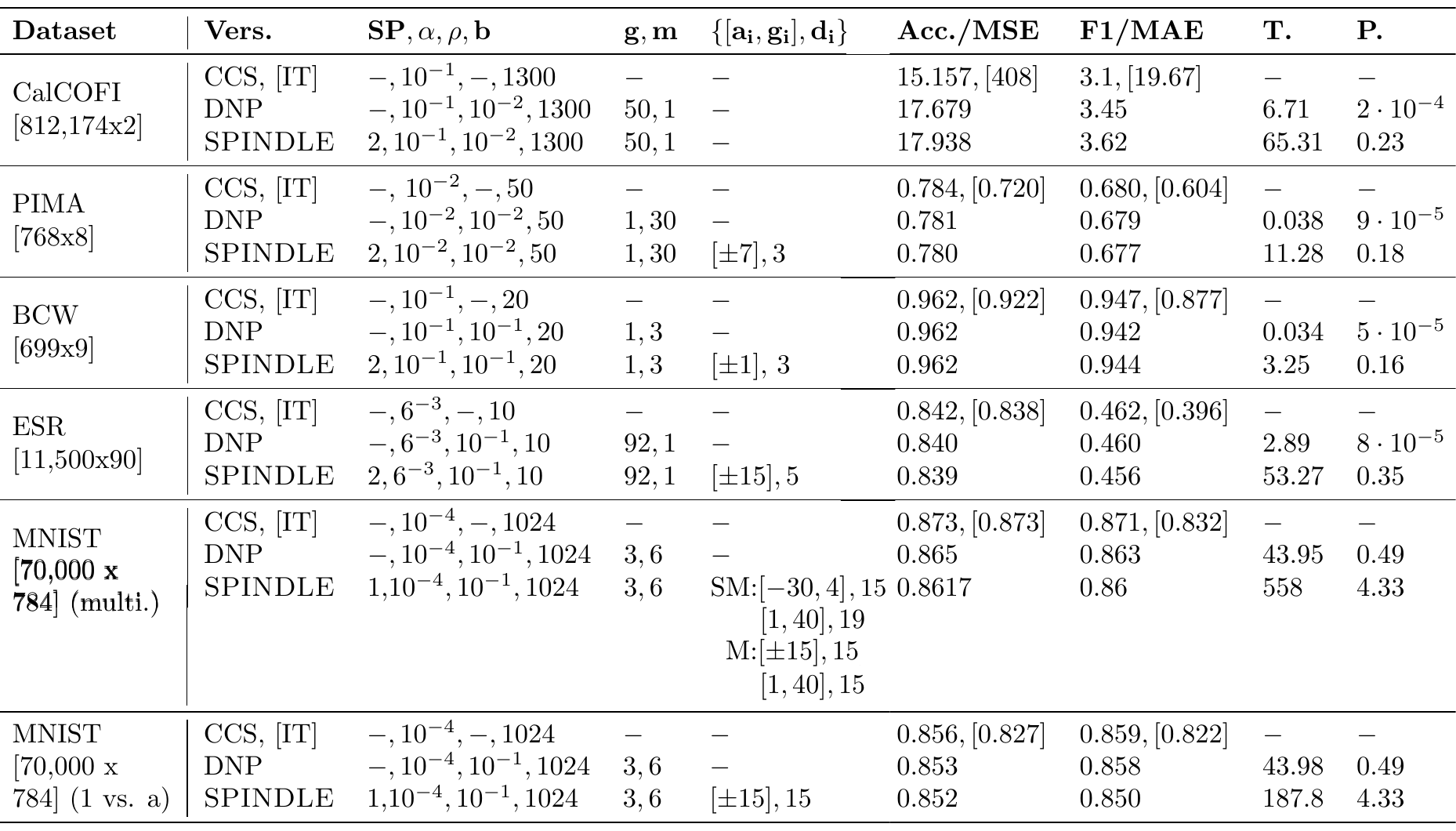}
		\vspace{-1.7em}
		\caption{Baseline Comparison with K-fold=5. Time to train (T.) and predict (P.) is in seconds.}
		\label{fig:AppbaselineComp}
	\end{subfigure}
	\caption{\centering \sys's Evaluation.}
	\label{figComparison}
\end{table*}

\subsection{Theoretical Analysis}

In Table \ref{fig:theoryAnalysis}, we provide the complete complexity analysis of \sys. $|ct|$ represents the maximum size of a ciphertext, i.e., the size of a fresh ciphertext at level $L$, $E$ and $D$ stand for encryption and decryption workloads. The DA packing approach incurs a higher computation complexity (with notably more plaintext-ciphertext multiplications and rescaling (M), and rotations (R)) but, as shown in Section~\ref{subsec:eval}, it is embarrassingly parallel, i.e., operations can be amortized by a factor $N_1 \cdot N_2$ (defined in Section~\ref{sec:MatrixMultSGD}) depending on the available threads.

\subsection{Evaluation Datasets}\label{app:datasets}

For linear regression, we use the CalCOFI dataset (with $n=812,174$ records and $c=2$ features)~\cite{CalCOFI}. It contains oceanographic data (e.g., salinity) that can be used to predict the water temperature. For logistic regression, we use three different datasets: (a) the Breast Cancer Wisconsin dataset (BCW, $n$ $=$ $699$, $c$ $=$ $9$)~\cite{BCW} contains patients' data that is employed to predict the presence of a breast cancer, (b) the PIMA dataset ($n$ $=$ $768$, $c$ $=$ $8$)~\cite{pima} contains medical observations collected from an Indian community that can be used to predict the presence of diabetes, and (c) the Epileptic Seizure Recognition dataset (ESR, $n$ $=$ $11,500$, $c$ $=$ $179$)~\cite{ESR} contains patients' data that can be used to predict a seizure. For multinomial regression, we test \sys on the MNIST dataset ($n$ $=$ $70,000$, $c$ $=$ $784$)~\cite{MNIST}, where the goal is to identify single-digits out of grey-scale images. We rely on these datasets to compare \sys with various baselines.

\subsection{Comparison with Prior Art}\label{app:compareprior}

In Table~\ref{fig:comparisonState}, we perform a qualitative and quantitative comparison of \sys with existing works. We consider a generic privacy-preserving centralized encrypted solution (CES), two distributed solutions, Drynx~\cite{Drynx} and Prio~\cite{corrigan2017prio}, which respectively rely on additive HE and secret sharing, and Helen~\cite{zheng2019helen}, a solution that employs a different distributed approach, the Alternating Direction Method of Multipliers (ADMM) proposed by Boyd et al.~\cite{boyd2011distributed}, to train regularized linear models.
CES represents a centralized solution, similar to existing works~\cite{kim2018secure,kim2018logistic,carpov2019privacy}, in which one DP outsources its encrypted data to a server that trains and evaluates a model. For a fair comparison, we estimate the execution time (without communication) of a generic centralized (outsourced) solution (CES) relying on the non-multiparty \textsc{ckks} scheme with security parameters that enable packing of the same number of values in one ciphertext. We use as a reference one of the most recent works on bootstrapping by Han and Dohyeong \cite{Han_BetterBootstrap}. 

\begin{table}[h]

	\centering
	\tiny
	\includegraphics[width=0.5\columnwidth]
	{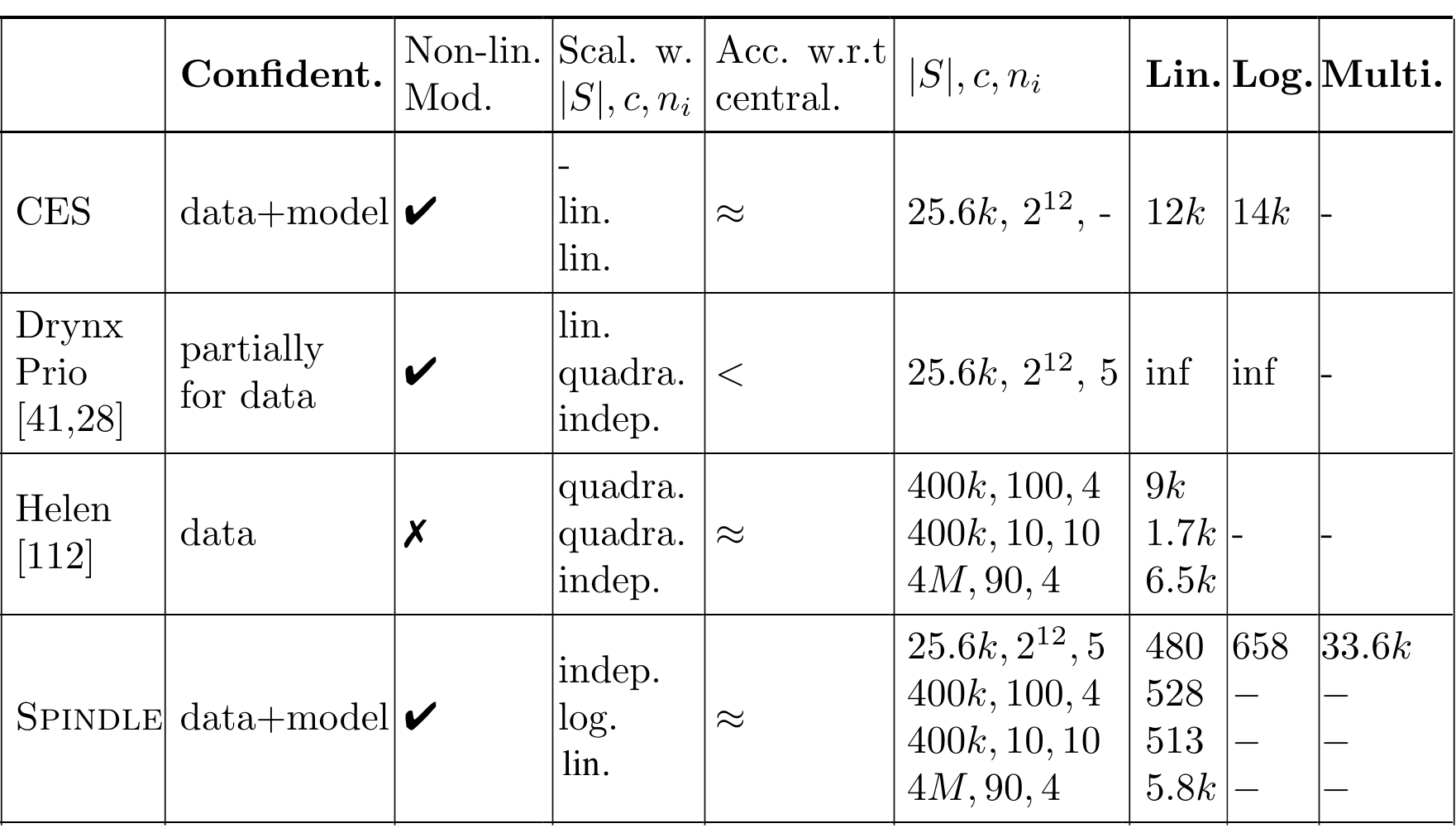}
	\caption{\small{Comparison with existing solutions. $|S|,\ c$ are the number of DPs and features, $n_i$ the size of each $DP_i$ local dataset. Timings for lin., log., multi. regressions training are in seconds.}}

	\label{fig:comparisonState}
\end{table}

In Helen, the DPs perform the ADMM optimization locally under a quantum-vulnerable additive HE cryptoscheme, and combine their results under secret-sharing. ADMM is less widespread than SGD, it is primarily designed for linear models and does not provide the same stability and convergence  guarantees than the cooperative gradient-descent \cite{zhang2015deep, Wang_CooperativeSGD,Wang_CooperativeSGD_pres,bottou2018optimization}, for which convergence can be derived from SGD. 
Since Helen's implementation is not available, we aim at providing an intuition on how it quantitatively compares with \sys. To this end, we used results reported in Helen~\cite{zheng2019helen} and performed similar experiments in \sys ((2),(3),(4) in Table \ref{fig:comparisonState}). 
We highlight here that the experiment environment is different, and these results provide only an idea of how these systems compare. For a fair comparison, we excluded the proof generation time in Helen and we notice that as \sys, Helen reported similar accuracy results as a non-secure centralized solution. We observe that \sys scales better than Helen when increasing $|S|$, as its execution time is almost the same in ((2) and (3)), and it also scales better with the number of features (almost 10x better in (4)).
\newpage
\section{Notations}\label{appendixNotation}

\begin{table}[h]
\centering
\small
\begin{tabular}{cl}
Symbol & Description \\ \midrule
\multicolumn{1}{c|}{$DP_i$}  & $i^{th}$ Data provider\\ \midrule
\multicolumn{1}{c|}{$S$, $|S|$}  & Set of DPs and its cardinality \\ \midrule
\multicolumn{1}{c|}{$\bm{X}_{n\times c}, \bm{y}_n$}  & Training dataset with $c$ features, \\ 
\multicolumn{1}{c|}{}  & $n$ samples, and label vector \\ \midrule
\multicolumn{1}{c|}{$(\bm{X}^{(i)}, \bm{y}^{(i)})$}  & $DP_i$'s part of the dataset \\ \midrule
\multicolumn{1}{c|}{$(\bm{X'},\cdot)$ \& $\bm{y'}$}  & Querier's evaluation data and prediction's output \\ \midrule
\multicolumn{1}{c|}{$cl$, $|cl|$}  & Set of class labels and its cardinality \\ \midrule
\multicolumn{1}{c|}{$\bm{X}[\phi,\cdot], \bm{X}[\cdot,\phi], \bm{y}[\phi]$}  & $\phi^{th}$ line and column of $\bm{X}$, element of vector $y$ \\ \midrule
\multicolumn{1}{c|}{$\bm{B} \in \bm{X}$}  & Random mini-batch of $b$ rows\\ \midrule
\multicolumn{1}{c|}{${\bm{W}^{(\cdot,j)}_G}$}  & Global model at iteration $j$ \\ \midrule
\multicolumn{1}{c|}{${\bm{W}^{(i,j,l)}}$}  & $DP_i$'s local model at global iter. $j$ and local iter. $l$ \\ \midrule
\multicolumn{1}{c|}{${\bm{W}^{(i,j)}}$}  & $DP_i$'s local model at global iter. $j$ \\ \midrule
\multicolumn{1}{c|}{${\bm{w}_G, \bm{w}}$}  & Vector of global and local weights\\ 
\multicolumn{1}{c|}{$\bm{w}^{(i,0)}$}  & Initial local weights of $DP_i$\\ \midrule
\multicolumn{1}{c|}{$lp$, $g,m$}  & Learning params., nbr. of global, local iterations \\ \midrule
\multicolumn{1}{c|}{$QR$}  & Querier request \\ \midrule
\multicolumn{1}{c|}{$\sigma(\cdot)$, $d$}  & Activation function, approx. degree \\ \midrule
\multicolumn{1}{c|}{ $a_m$}  & Multiplicative depth of $\sigma(\cdot)$ \\ \midrule
\multicolumn{1}{c|}{$I(\cdot)$}  & Indicator function \\ \midrule
\multicolumn{1}{c|}{ $n_{i}$}  & The number of data samples per $DP_i$ \\ \midrule
\multicolumn{1}{c|}{ $P^2(x)$}  & Next power of 2 of $x$ \\ \midrule
\multicolumn{1}{c|}{ $N_1, N_2$}  & Diagonal approach parameters.\\
\multicolumn{1}{c|}{$\alpha$, $\rho$}  & Learning and elastic rates \\ \toprule\toprule
\multicolumn{1}{c|}{$sk, pk$, $evk$}  & Secret, public, evaluation keys  \\ \midrule
\multicolumn{1}{c|}{ $\langle \bm{v} \rangle$}  &  Encrypted vector $\bm{v}$ \\ \midrule
\multicolumn{1}{c|}{$|ct|$}  & Size of fresh ciphertext $ct$ \\ \midrule
\multicolumn{1}{c|}{$N$, $Q$}  & Ring dimension, Fresh ciphertext modulus \\ \midrule
\multicolumn{1}{c|}{$L$}  & Number of available levels \\ \midrule
\multicolumn{1}{c|}{$\tau_b$}  & Minimum level for $\text{Dbootstrap}(\cdot)$ \\ \midrule
\multicolumn{1}{c|}{$\eta$}  & Std. deviation of the noise distribution \\ \midrule
\multicolumn{1}{c|}{$\Delta$, $mc$}  & Plaintext Scale, Chain of moduli variables \\ \midrule
\multicolumn{1}{c|}{$\text{DM}$}  & Dot product \\ \bottomrule
\end{tabular}
\caption{Frequently Used Symbols and Notations.}
\vspace{-3.5em}
\label{TableNotations}
\end{table}

%% file: main.bbl
\begin{thebibliography}{100}

\bibitem{tensorflow}
M.~Abadi et~al.
\newblock {TensorFlow}: Large-scale machine learning on heterogeneous systems,
  2015.
\newblock Software available from tensorflow.org.

\bibitem{abadi2016deep}
M.~Abadi et~al.
\newblock Deep learning with differential privacy.
\newblock In {\em ACM CCS}, 2016.

\bibitem{Akavia_WAHC}
A.~Akavia, H.~Shaul, M.~Weiss, and Z.~Yakhini.
\newblock Linear-{R}egression on {P}acked {E}ncrypted {D}ata in the
  {T}wo-{S}erver {M}odel.
\newblock In {\em ACM WAHC}, 2019.

\bibitem{HEStandardPaper}
M.~Albrecht et~al.
\newblock Homomorphic {E}ncryption {S}ecurity {S}tandard.
\newblock Technical report, HomomorphicEncryption.org, 2018.

\bibitem{Albrecht2015OnTC}
M.~R. Albrecht, R.~Player, and S.~Scott.
\newblock On the concrete hardness of learning with errors.
\newblock {\em J. of Mathematical Cryptology}, 2015.

\bibitem{Algesheimer2002}
S.~V. Algesheimer~J., Camenisch~J.
\newblock Efficient computation modulo a shared secret with application to the
  generation of shared safe-prime products.
\newblock In {\em CRYPTO}, 2002.

\bibitem{aono2016scalable}
Y.~Aono, T.~Hayashi, L.~Trieu~Phong, and L.~Wang.
\newblock Scalable and secure logistic regression via homomorphic encryption.
\newblock In {\em ACM CODASPY}, 2016.

\bibitem{baum2016efficient}
C.~Baum, I.~Damg{\aa}rd, S.~Oechsner, and C.~Peikert.
\newblock Efficient commitments and zero-knowledge protocols from ring-sis with
  applications to lattice-based threshold cryptosystems.
\newblock {\em IACR Cryptol. ePrint Arch.}, 2016.

\bibitem{baum2020concretely}
C.~Baum and A.~Nof.
\newblock Concretely-efficient zero-knowledge arguments for arithmetic circuits
  and their application to lattice-based cryptography.
\newblock In {\em PKC}, 2020.

\bibitem{BCW}
Breast {C}ancer {W}isconsin ({O}riginal).
\newblock
  \url{https://archive.ics.uci.edu/ml/datasets/breast+cancer+wisconsin+(original)}
  (14.02.2020).

\bibitem{beam2018big}
A.~L. Beam and I.~S. Kohane.
\newblock Big data and machine learning in health care.
\newblock {\em Jama}, 2018.

\bibitem{boemer2019ngraph}
F.~Boemer, A.~Costache, R.~Cammarota, and C.~Wierzynski.
\newblock n{G}raph-{HE}2: {A} {H}igh-{T}hroughput {F}ramework for {N}eural
  {N}etwork {I}nference on {E}ncrypted {D}ata.
\newblock In {\em ACM WAHC}, 2019.

\bibitem{bogdanov2016rmind}
D.~Bogdanov, L.~Kamm, S.~Laur, and V.~Sokk.
\newblock Rmind: a tool for cryptographically secure statistical analysis.
\newblock {\em IEEE TDSC}, 2016.

\bibitem{Bonawitz2019}
K.~Bonawitz et~al.
\newblock Towards federated learning at scale: System design.
\newblock In {\em SysML}, 2019.

\bibitem{bonte2018privacy}
C.~Bonte and F.~Vercauteren.
\newblock Privacy-preserving logistic regression training.
\newblock {\em BMC medical genomics}, 2018.

\bibitem{bos2013improved}
J.~W. Bos, K.~Lauter, J.~Loftus, and M.~Naehrig.
\newblock Improved security for a ring-based fully homomorphic encryption
  scheme.
\newblock In {\em IMACC}, 2013.

\bibitem{bost2015machine}
R.~Bost, R.~A. Popa, S.~Tu, and S.~Goldwasser.
\newblock Machine learning classification over encrypted data.
\newblock In {\em NDSS}, 2015.

\bibitem{bottou2018optimization}
L.~Bottou, F.~E. Curtis, and J.~Nocedal.
\newblock Optimization methods for large-scale machine learning.
\newblock {\em Siam Review}, 2018.

\bibitem{boyd2011distributed}
S.~Boyd, N.~Parikh, E.~Chu, B.~Peleato, J.~Eckstein, et~al.
\newblock Distributed optimization and statistical learning via the alternating
  direction method of multipliers.
\newblock {\em Foundations and Trends in Machine learning}, 2011.

\bibitem{CalCOFI}
Cal{COFI}, over 60 years of oceanographic data.
\newblock \url{https://www.kaggle.com/sohier/calcofi} (05.03.2020).

\bibitem{carpov2019privacy}
S.~Carpov, N.~Gama, M.~Georgieva, and J.~R. Troncoso-Pastoriza.
\newblock Privacy-preserving semi-parallel logistic regression training with
  fully homomorphic encryption.
\newblock {\em IACR Cryptology ePrint Archive}, 2019.

\bibitem{chaudhuri2009privacy}
K.~Chaudhuri and C.~Monteleoni.
\newblock Privacy-preserving logistic regression.
\newblock In {\em NIPS}, 2009.

\bibitem{chen2018logistic}
H.~Chen, R.~Gilad-Bachrach, K.~Han, Z.~Huang, A.~Jalali, K.~Laine, and
  K.~Lauter.
\newblock Logistic regression over encrypted data from fully homomorphic
  encryption.
\newblock {\em BMC medical genomics}, 2018.

\bibitem{cheon2019hybrid}
J.~H. Cheon, M.~Hhan, S.~Hong, and Y.~Son.
\newblock A hybrid of dual and meet-in-the-middle attack on sparse and ternary
  secret {LWE}.
\newblock {\em IEEE Access}, 2019.

\bibitem{cheon2017homomorphic}
J.~H. Cheon, A.~Kim, M.~Kim, and Y.~Song.
\newblock Homomorphic encryption for arithmetic of approximate numbers.
\newblock In {\em ASIACRYPT}, 2017.

\bibitem{Cho_GWAS}
H.~Cho, D.~Wu, and B.~Berger.
\newblock Secure genome-wide association analysis using multiparty computation.
\newblock {\em Nature Biotech.}, 2018.

\bibitem{chu2007map}
C.-T. Chu et~al.
\newblock Map-reduce for machine learning on multicore.
\newblock In {\em NIPS}, 2007.

\bibitem{corrigan2017prio}
H.~Corrigan-Gibbs and D.~Boneh.
\newblock Prio: {P}rivate, {R}obust, and {C}omputation of {A}ggregate
  {S}tatistics.
\newblock In {\em NSDI}, 2017.

\bibitem{crawford2018doing}
J.~L. Crawford, C.~Gentry, S.~Halevi, D.~Platt, and V.~Shoup.
\newblock Doing real work with {FHE}: The case of logistic regression.
\newblock In {\em ACM WAHC}, 2018.

\bibitem{damgaard2012multiparty}
I.~Damg{\aa}rd, V.~Pastro, N.~Smart, and S.~Zakarias.
\newblock Multiparty computation from somewhat homomorphic encryption.
\newblock In {\em CRYPTO}, 2012.

\bibitem{dean2008mapreduce}
J.~Dean and S.~Ghemawat.
\newblock Map{R}educe: simplified data processing on large clusters.
\newblock {\em Communications of the ACM}, 2008.

\bibitem{NN_SGD}
S.~S. Du, J.~D. Lee, H.~Li, L.~Wang, and X.~Zhai.
\newblock Gradient descent finds global minima of deep neural networks.
\newblock {\em CoRR}, abs/1811.03804, 2018.

\bibitem{du2018gradient}
S.~S. Du, X.~Zhai, B.~Poczos, and A.~Singh.
\newblock Gradient descent provably optimizes over-parameterized neural
  networks.
\newblock {\em arXiv preprint arXiv:1810.02054}, 2018.

\bibitem{du2018privacy}
W.~Du, A.~Li, and Q.~Li.
\newblock Privacy-{P}reserving {M}ultiparty {L}earning {F}or {L}ogistic
  {R}egression.
\newblock In {\em SecureComm}, 2018.

\bibitem{elgamal1985public}
T.~ElGamal.
\newblock A public key cryptosystem and a signature scheme based on discrete
  logarithms.
\newblock {\em IEEE Trans-IT}, 1985.

\bibitem{erickson2017machine}
B.~J. Erickson, P.~Korfiatis, Z.~Akkus, and T.~L. Kline.
\newblock Machine learning for medical imaging.
\newblock {\em Radiographics}, 2017.

\bibitem{ESR}
Epileptic {S}eizure {R}ecognition {D}ataset.
\newblock
  \url{https://archive.ics.uci.edu/ml/datasets/Epileptic+Seizure+Recognition}
  (14.02.2020).

\bibitem{fan2012somewhat}
J.~Fan and F.~Vercauteren.
\newblock Somewhat practical fully homomorphic encryption.
\newblock {\em IACR Cryptology ePrint Archive}, 2012.

\bibitem{fredrikson2015model}
M.~Fredrikson, S.~Jha, and T.~Ristenpart.
\newblock Model inversion attacks that exploit confidence information and basic
  countermeasures.
\newblock In {\em ACM CCS}, 2015.

\bibitem{Unlynx}
D.~Froelicher, P.~Egger, J.~S. Sousa, J.~L. Raisaro, Z.~Huang, C.~V. Mouchet,
  B.~Ford, and J.-P. Hubaux.
\newblock Unlynx: A decentralized system for privacy-conscious data sharing.
\newblock {\em PETS}, 2017.

\bibitem{Drynx}
D.~{Froelicher}, J.~R. {Troncoso-Pastoriza}, J.~S. {Sousa}, and J.~{Hubaux}.
\newblock Drynx: Decentralized, secure, verifiable system for statistical
  queries and machine learning on distributed datasets.
\newblock {\em IEEE TIFS}, 2020.

\bibitem{gascon2017privacy}
A.~Gasc{\'o}n, P.~Schoppmann, B.~Balle, M.~Raykova, J.~Doerner, S.~Zahur, and
  D.~Evans.
\newblock Privacy-preserving distributed linear regression on high-dimensional
  data.
\newblock {\em PETS}, 2017.

\bibitem{GDPR}
{The {EU} General Data Protection Regulation}.
\newblock \url{https://eugdpr.org/ (10.11.2019)}.

\bibitem{giacomelli2018privacy}
I.~Giacomelli, S.~Jha, M.~Joye, C.~D. Page, and K.~Yoon.
\newblock Privacy-preserving ridge regression with only linearly-homomorphic
  encryption.
\newblock In {\em ACNS}, 2018.

\bibitem{gilad2016cryptonets}
R.~Gilad-Bachrach, N.~Dowlin, K.~Laine, K.~Lauter, M.~Naehrig, and J.~Wernsing.
\newblock Cryptonets: Applying neural networks to encrypted data with high
  throughput and accuracy.
\newblock In {\em ICML}, 2016.

\bibitem{Go}
{Go Programming Language}.
\newblock \url{https://golang.org, (10.11.2019)}.

\bibitem{Quantum1}
L.~{Gomes}.
\newblock Quantum computing: Both here and not here.
\newblock {\em IEEE Spectrum}, 2018.

\bibitem{NeuralBook}
I.~Goodfellow, Y.~Bengio, and A.~Courville.
\newblock {\em Deep Learning}.
\newblock MIT Press, 2016.
\newblock \url{http://www.deeplearningbook.org}.

\bibitem{gooquantumsuprem}
{Google {CEO} Sundar Pichai on achieving quantum supremacy}.
\newblock \url{https://tinyurl.com/y5rnowlc} (07.11.2019).

\bibitem{graepel2012ml}
T.~Graepel, K.~Lauter, and M.~Naehrig.
\newblock {ML} confidential: Machine learning on encrypted data.
\newblock In {\em ICISC}, 2012.

\bibitem{Helib}
S.~Halevi and V.~Shoup.
\newblock Algorithms in {HE}lib.
\newblock In {\em CRYPTO}, 2014.

\bibitem{Han_BetterBootstrap}
K.~Han and D.~Ki.
\newblock Better bootstrapping for approximate homomorphic encryption.
\newblock In {\em CT-RSA}, 2020.

\bibitem{hesamifard2018privacy}
E.~Hesamifard, H.~Takabi, M.~Ghasemi, and R.~N. Wright.
\newblock Privacy-preserving machine learning as a service.
\newblock {\em PETS}, 2018.

\bibitem{hitaj2017deep}
B.~Hitaj, G.~Ateniese, and F.~Perez-Cruz.
\newblock Deep models under the {GAN}: information leakage from collaborative
  deep learning.
\newblock In {\em ACM CCS}, 2017.

\bibitem{huang2019dp}
Z.~Huang, R.~Hu, Y.~Guo, E.~Chan-Tin, and Y.~Gong.
\newblock {DP-ADMM}: {ADMM}-based distributed learning with differential
  privacy.
\newblock {\em IEEE TIFS}, 2019.

\bibitem{IBMQuantum}
{Quantum Computing is “no longer science fiction,” says IBM}.
\newblock \url{https://tinyurl.com/y4zvlsll}, (10.02.2020).

\bibitem{Jagadeesh692}
K.~A. Jagadeesh, D.~J. Wu, J.~A. Birgmeier, D.~Boneh, and G.~Bejerano.
\newblock Deriving genomic diagnoses without revealing patient genomes.
\newblock {\em Science}, 2017.

\bibitem{jayaraman2019evaluating}
B.~Jayaraman and D.~Evans.
\newblock Evaluating differentially private machine learning in practice.
\newblock In {\em USENIX Security}, 2019.

\bibitem{jayaraman2018distributed}
B.~Jayaraman, L.~Wang, D.~Evans, and Q.~Gu.
\newblock Distributed learning without distress: Privacy-preserving empirical
  risk minimization.
\newblock In {\em NIPS}, 2018.

\bibitem{jiang2019securelr}
Y.~Jiang et~al.
\newblock Secure{LR}: Secure logistic regression model via a hybrid
  cryptographic protocol.
\newblock {\em IEEE TCB}, 2019.

\bibitem{juvekar2018gazelle}
C.~Juvekar, V.~Vaikuntanathan, and A.~Chandrakasan.
\newblock {GAZELLE}: A low latency framework for secure neural network
  inference.
\newblock In {\em USENIX Security}, 2018.

\bibitem{ndas2}
Why we shouldn’t disregard the nda.
\newblock \url{tinyurl.com/y4hdr42d}.

\bibitem{kim2018logistic}
A.~Kim, Y.~Song, M.~Kim, K.~Lee, and J.~H. Cheon.
\newblock Logistic regression model training based on the approximate
  homomorphic encryption.
\newblock {\em BMC genomics}, 2018.

\bibitem{kim2020}
M.~{Kim}, J.~{Lee}, L.~{Ohno-Machado}, and X.~{Jiang}.
\newblock Secure and differentially private logistic regression for
  horizontally distributed data.
\newblock {\em IEEE TIFS}, 2020.

\bibitem{kim2018secure}
M.~Kim, Y.~Song, S.~Wang, Y.~Xia, and X.~Jiang.
\newblock Secure logistic regression based on homomorphic encryption: Design
  and evaluation.
\newblock {\em JMIR medical informatics}, 2018.

\bibitem{Jaggi_2019}
A.~Koloskova, S.~U. Stich, and M.~Jaggi.
\newblock Decentralized stochastic optimization and gossip algorithms with
  compressed communication.
\newblock {\em CoRR}, abs/1902.00340, 2019.

\bibitem{Konency2016fed}
J.~Konečný, H.~McMahan, D.~Ramage, and P.~Richtárik.
\newblock Federated optimization: Distributed machine learning for on-device
  intelligence.
\newblock {\em arXiv preprint arXiv:1610.02527}, 2016.

\bibitem{Kumar_2015}
A.~Kumar, J.~Naughton, and J.~M. Patel.
\newblock Learning generalized linear models over normalized data.
\newblock In {\em ACM SIGMOD}, 2015.

\bibitem{MNIST}
Y.~LeCun and C.~Cortes.
\newblock {MNIST} handwritten digit database.
\newblock {\em \url{http://yann.lecun.com/exdb/mnist/}}, 2010.

\bibitem{leung2015machine}
M.~K. Leung, A.~Delong, B.~Alipanahi, and B.~J. Frey.
\newblock Machine learning in genomic medicine: a review of computational
  problems and data sets.
\newblock {\em Proceedings of the IEEE}, 2015.

\bibitem{Nvidia_Fed}
W.~Li et~al.
\newblock Privacy-preserving federated brain tumour segmentation.
\newblock In {\em MLMI}, 2019.

\bibitem{libert2018lattice}
B.~Libert, S.~Ling, K.~Nguyen, and H.~Wang.
\newblock Lattice-based zero-knowledge arguments for integer relations.
\newblock In {\em CRYPTO}, 2018.

\bibitem{lindell2017simulate}
Y.~Lindell.
\newblock How to simulate it--a tutorial on the simulation proof technique.
\newblock In {\em Tutorials on the Foundations of Cryptography}. 2017.

\bibitem{Lindner2011}
R.~Lindner and C.~Peikert.
\newblock Better key sizes (and attacks) for {LWE}-based encryption.
\newblock In {\em CT-RSA}, 2011.

\bibitem{glm_applications}
J.~K. Lindsey.
\newblock {\em Applying generalized linear models}.
\newblock Springer Science \& Business Media, 2000.

\bibitem{ndas1}
Why {NDA}s often don't work when expected to do so and what to do about it.
\newblock \url{https://tinyurl.com/y64qlzs9}.

\bibitem{liu2017oblivious}
J.~Liu, M.~Juuti, Y.~Lu, and N.~Asokan.
\newblock Oblivious neural network predictions via minionn transformations.
\newblock In {\em ACM CCS}, 2017.

\bibitem{lyubashevsky2020practical}
V.~Lyubashevsky, N.~K. Nguyen, and G.~Seiler.
\newblock Practical lattice-based zero-knowledge proofs for integer relations.
\newblock In {\em ACM CCS}, 2020.

\bibitem{lyubashevsky2010ideal}
V.~Lyubashevsky, C.~Peikert, and O.~Regev.
\newblock On ideal lattices and learning with errors over rings.
\newblock In {\em EUROCRYPT}, 2010.

\bibitem{mcmahan2016communication}
H.~B. McMahan, E.~Moore, D.~Ramage, S.~Hampson, et~al.
\newblock Communication-efficient learning of deep networks from decentralized
  data.
\newblock {\em arXiv preprint arXiv:1602.05629}, 2016.

\bibitem{federatedLearning1}
H.~B. McMahan, E.~Moore, D.~Ramage, and B.~A. y~Arcas.
\newblock Federated learning of deep networks using model averaging.
\newblock {\em CoRR}, abs/1602.05629, 2016.

\bibitem{McMahan2018}
H.~B. McMahan, D.~Ramage, K.~Talwar, and L.~Zhang.
\newblock Learning differentially private recurrent language models.
\newblock In {\em ICLR}, 2018.

\bibitem{Melis2019}
L.~{Melis}, C.~{Song}, E.~{De Cristofaro}, and V.~{Shmatikov}.
\newblock Exploiting unintended feature leakage in collaborative learning.
\newblock In {\em IEEE S\&P}, 2019.

\bibitem{lattigo}
Lattigo: A library for lattice-based homomorphic encryption in go.
\newblock \url{https://github.com/ldsec/lattigo} (14.02.2019).

\bibitem{mininet}
Mininet.
\newblock \url{http://mininet.org} (13.12.2019).

\bibitem{mohassel2018aby}
P.~Mohassel and P.~Rindal.
\newblock {ABY} 3: a mixed protocol framework for machine learning.
\newblock In {\em ACM CCS}, 2018.

\bibitem{mohassel2017secureml}
P.~Mohassel and Y.~Zhang.
\newblock Secure{ML}: A system for scalable privacy-preserving machine
  learning.
\newblock In {\em IEEE S\&P}, 2017.

\bibitem{Mosca_Quantum}
M.~{Mosca}.
\newblock Cybersecurity in an era with quantum computers: Will we be ready?
\newblock {\em IEEE S\&P}, 2018.

\bibitem{BigDataMedical}
M.~Mostert, A.~Bredenoord, M.~Biesaart, and J.~Delden.
\newblock Big data in medical research and {EU} data protection law: challenges
  to the consent or anonymise approach.
\newblock {\em European Journal of Human Genetics}, 2016.

\bibitem{mouchet2019distributedbfv}
C.~Mouchet, J.~R. Troncoso-pastoriza, J.-P. Bossuat, and J.~P. Hubaux.
\newblock Multiparty homomorphic encryption: From theory to practice.
\newblock In {\em Tech. Report \url{https://eprint.iacr.org/2020/304}}, 2019.

\bibitem{Nasr2019}
M.~{Nasr}, R.~{Shokri}, and A.~{Houmansadr}.
\newblock Comprehensive privacy analysis of deep learning: Passive and active
  white-box inference attacks against centralized and federated learning.
\newblock In {\em IEEE S\&P}, 2019.

\bibitem{Quantum4}
C.~Neill et~al.
\newblock A blueprint for demonstrating quantum supremacy with superconducting
  qubits.
\newblock {\em Science}, 2018.

\bibitem{GLM}
J.~A. Nelder and R.~W.~M. Wedderburn.
\newblock Generalized linear models.
\newblock {\em Journal of the Royal Statistical Society}, 1972.

\bibitem{nesterov2005smooth}
Y.~Nesterov.
\newblock Smooth minimization of non-smooth functions.
\newblock {\em Mathematical programming}, 2005.

\bibitem{nikolaenko2013privacy}
V.~Nikolaenko, U.~Weinsberg, S.~Ioannidis, M.~Joye, D.~Boneh, and N.~Taft.
\newblock Privacy-preserving ridge regression on hundreds of millions of
  records.
\newblock In {\em IEEE S\&P}, 2013.

\bibitem{Onet}
{Cothority network library}.
\newblock \url{https://github.com/dedis/onet,}.

\bibitem{paillier1999public}
P.~Paillier.
\newblock Public-key cryptosystems based on composite degree residuosity
  classes.
\newblock In {\em EUROCRYPT}, 1999.

\bibitem{pytorch}
A.~Paszke et~al.
\newblock Automatic differentiation in {P}y{T}orch.
\newblock 2017.

\bibitem{pathak2010multiparty}
M.~Pathak, S.~Rane, and B.~Raj.
\newblock Multiparty differential privacy via aggregation of locally trained
  classifiers.
\newblock In {\em NIPS}, 2010.

\bibitem{8241854}
L.~T. {Phong}, Y.~{Aono}, T.~{Hayashi}, L.~{Wang}, and S.~{Moriai}.
\newblock Privacy-preserving deep learning via additively homomorphic
  encryption.
\newblock {\em IEEE TIFS}, 2018.

\bibitem{pima}
Pima {I}ndians {D}iabetes {D}ataset.
\newblock \url{https://tinyurl.com/y8o3x8me} (14.04.2018).

\bibitem{delphi}
M.~Pratyush, R.~Lehmkuhl, A.~Srinivasan, W.~Zheng, and R.~A. Popa.
\newblock Delphi: A cryptographic inference service for neural networks.
\newblock In {\em USENIX Security}, 2020.

\bibitem{rachuri2019trident}
R.~Rachuri and A.~Suresh.
\newblock Trident: Efficient 4{PC} framework for privacy preserving machine
  learning.
\newblock In {\em NDSS}, 2020.

\bibitem{riazi2018chameleon}
M.~S. Riazi et~al.
\newblock Chameleon: A hybrid secure computation framework for machine learning
  applications.
\newblock In {\em ASIACCS}, 2018.

\bibitem{riazi2019xonn}
M.~S. Riazi, M.~Samragh, H.~Chen, K.~Laine, K.~E. Lauter, and F.~Koushanfar.
\newblock {XONN}: {XNOR}-based oblivious deep neural network inference.
\newblock In {\em USENIX Security}, 2019.

\bibitem{rouhani2018deepsecure}
B.~D. Rouhani, M.~S. Riazi, and F.~Koushanfar.
\newblock Deepsecure: Scalable provably-secure deep learning.
\newblock In {\em ACM DAC}, 2018.

\bibitem{Schoenmakers2006}
B.~Schoenmakers and P.~Tuyls.
\newblock Efficient computation modulo a shared secret with application to the
  generation of shared safe-prime products.
\newblock In {\em EUROCRYPT}, 2006.

\bibitem{schoppmann2019make}
P.~Schoppmann, A.~Gascon, M.~Raykova, and B.~Pinkas.
\newblock Make some room for the zeros: Data sparsity in secure distributed
  machine learning.
\newblock In {\em ACM CCS}, 2019.

\bibitem{scikit-learn}
{Scikit-learn, Machine Learning in {P}ython}.
\newblock \url{https://scikit-learn.org/stable/, (29.02.2020)}.

\bibitem{shamir1979share}
A.~Shamir.
\newblock How to share a secret.
\newblock {\em Communications of the ACM}, 1979.

\bibitem{shokri2015privacy}
R.~Shokri and V.~Shmatikov.
\newblock Privacy-preserving deep learning.
\newblock In {\em ACM CCS}, 2015.

\bibitem{shokri2017membership}
R.~Shokri, M.~Stronati, C.~Song, and V.~Shmatikov.
\newblock Membership inference attacks against machine learning models.
\newblock In {\em IEEE S\&P}, 2017.

\bibitem{AI_competition}
I.~Stoica, D.~Song, R.~A. Popa, D.~Patterson, M.~W. Mahoney, R.~Katz, A.~D.
  Joseph, M.~Jordan, J.~M. Hellerstein, J.~E. Gonzalez, et~al.
\newblock A berkeley view of systems challenges for ai.
\newblock {\em arXiv preprint arXiv:1712.05855}, 2017.

\bibitem{Quantum3}
B.~Terhal.
\newblock Quantum supremacy, here we come.
\newblock {\em Nature Physics}, 2018.

\bibitem{BigDataIssues}
R.~Toshniwal, K.~Dastidar, and A.~Nath.
\newblock Big data security issues and challenges.
\newblock {\em International Journal of Innovative Research in Advanced
  Engineering}, 2015.

\bibitem{toulis2014statistical}
P.~Toulis, E.~Airoldi, and J.~Rennie.
\newblock Statistical analysis of stochastic gradient methods for generalized
  linear models.
\newblock In {\em ICML}, 2014.

\bibitem{truex2019hybrid}
S.~Truex et~al.
\newblock A hybrid approach to privacy-preserving federated learning.
\newblock In {\em ACM AISec}, 2019.

\bibitem{verbraeken2019survey}
J.~Verbraeken, M.~Wolting, J.~Katzy, J.~Kloppenburg, T.~Verbelen, and J.~S.
  Rellermeyer.
\newblock A survey on distributed machine learning.
\newblock {\em arXiv preprint arXiv:1912.09789}, 2019.

\bibitem{wagh2019securenn}
S.~Wagh, D.~Gupta, and N.~Chandran.
\newblock Secure{NN}: 3-party secure computation for neural network training.
\newblock {\em PETS}, 2019.

\bibitem{Wang_CooperativeSGD}
J.~Wang and G.~Joshi.
\newblock Cooperative {SGD:} {A} unified framework for the design and analysis
  of communication-efficient {SGD} algorithms.
\newblock {\em CoRR}, abs/1808.07576, 2018.

\bibitem{Wang_CooperativeSGD_pres}
J.~Wang and G.~Joshi.
\newblock Cooperative {SGD}: A unified framework for the design and analysis of
  communication-efficient sgd algorithms.
\newblock In {\em ICML CodML Workshop}, 2019.

\bibitem{Wang2019}
Z.~{Wang}, M.~{Song}, Z.~{Zhang}, Y.~{Song}, Q.~{Wang}, and H.~{Qi}.
\newblock Beyond inferring class representatives: User-level privacy leakage
  from federated learning.
\newblock In {\em IEEE INFOCOM}, 2019.

\bibitem{Wolinsky2012ScalableAG}
D.~I. Wolinsky, H.~Corrigan-Gibbs, B.~Ford, and A.~Johnson.
\newblock Scalable anonymous group communication in the anytrust model.
\newblock Technical report, Naval Research Lab Washington DC, 2012.

\bibitem{yang2019efficient}
R.~Yang, M.~H. Au, Z.~Zhang, Q.~Xu, Z.~Yu, and W.~Whyte.
\newblock Efficient lattice-based zero-knowledge arguments with standard
  soundness: construction and applications.
\newblock In {\em CRYPTO}, 2019.

\bibitem{yao1986generate}
A.~C.-C. Yao.
\newblock How to generate and exchange secrets.
\newblock In {\em IEEE SFCS}, 1986.

\bibitem{revisitinghybridattackssparsekeys}
{Yongha Son and Jung Hee Cheon}.
\newblock Revisiting the hybrid attack on sparse and ternary secret {LWE}.
\newblock {\em Technical Report \url{https://eprint.iacr.org/2019/1019},},
  2019.

\bibitem{Quantum2_GoogleAI}
A.~Zalcman et~al.
\newblock Quantum supremacy using a programmable superconducting processor.
\newblock {\em Nature}, 2019.

\bibitem{Zhang_BigDataSecurity}
D.~Zhang.
\newblock Big data security and privacy protection.
\newblock In {\em ICMCS}, 2018.

\bibitem{zhang2015deep}
S.~Zhang, A.~E. Choromanska, and Y.~LeCun.
\newblock Deep learning with elastic averaging sgd.
\newblock In {\em NIPS}, 2015.

\bibitem{Zhang_2004}
T.~Zhang.
\newblock Solving large scale linear prediction problems using stochastic
  gradient descent algorithms.
\newblock In {\em ICML}, 2004.

\bibitem{zheng2019helen}
W.~Zheng, R.~A. Popa, J.~E. Gonzalez, and I.~Stoica.
\newblock Helen: {M}aliciously {S}ecure {C}oopetitive {L}earning for {L}inear
  {M}odels.
\newblock In {\em IEEE S\&P}, 2019.

\bibitem{NIPS2019_9617}
L.~Zhu, Z.~Liu, and S.~Han.
\newblock Deep leakage from gradients.
\newblock In {\em NIPS}. 2019.

\bibitem{Zhu2016}
X.~Zhu, C.~Vondrick, C.~C. Fowlkes, and D.~Ramanan.
\newblock Do we need more training data?
\newblock {\em Int. J. Comput. Vision}, 2016.

\end{thebibliography}
